# On Separation, Randomness and Linearity for Network Codes over Finite Fields [*]

Siddharth Ray [†]   Michelle Effros [‡]   Muriel Médard [†]   Ralf Koetter [§]
Tracey Ho [‡]   David Karger [¶]   Jinane Abounadi [†]

## Abstract

We examine the issue of separation and code design for networks that operate over finite fields. We demonstrate that source-channel (or source-network) separation holds for several canonical network examples like the noisy multiple access channel and the erasure degraded broadcast channel, when the whole network operates over a common finite field. This robustness of separation is predicated on the fact that noise and inputs are independent, and we examine the failure of separation when noise is dependent on inputs in multiple access channels.

Our approach is based on the sufficiency of linear codes. Using a simple and unifying framework, we not only re-establish with economy the optimality of linear codes for single-transmitter, single-receiver channels and for Slepian-Wolf source coding, but also establish the optimality of linear codes for multiple access and for erasure degraded broadcast channels. The linearity allows us to obtain simple optimal code constructions and to study capacity regions of the noisy multiple access and the degraded broadcast channel. The linearity of both source and network coding blurs the delineation between source and network codes. While our results point to the fact that separation of source coding and channel coding is optimal in some canonical networks, we show that decomposing networks into canonical subnetworks may not be effective. Thus, we argue that it may be the lack of decomposability of a network into canonical network modules, rather than the lack of separation between source and channel coding, that presents major challenges for coding over networks.

[*]The material in this paper was presented in part at the Communication Theory Symposium, Globecom 2003, San Francisco, California, December 2003; $41^{st}$ Annual Allerton Conference on Comunication, Control and Computing, Allerton, Illinois, October 2003 and DIMACS workshop on Network Information Theory, 2003.

[†]S. Ray (sray@mit.edu), M. Médard (medard@mit.edu) and J. Abounadi (jinane@mit.edu) are with the Laboratory for Information and Decision Systems (LIDS), Massachusetts Institute of Technology, Cambridge, MA 02139. Work is supported by NSF grant CCR-0220039, University of Illinois subaward 02-194, Hewlett-Packard 008542-008 and MURI 6893790.

[‡]M. Effros (effros@caltech.edu) and T. Ho (tho@caltech.edu) are with the Department of Electrical Engineering, 136-93, California Institute of Technology, Pasadena, CA 91125. Work is supported by Caltech's Lee Center for Advanced Networking.

[§]R. Koetter (koetter@uiuc.edu) is with the Coordinated Science Laboratory, University of Illinois at Urbana-Champaign, Urbana, IL 61801.

[¶]D. Karger (karger@mit.edu) is with the Computer Science and Artificial Intelligence Laboratory (CSAIL), Massachusetts Institute of Technology, Cambridge, MA 02139.

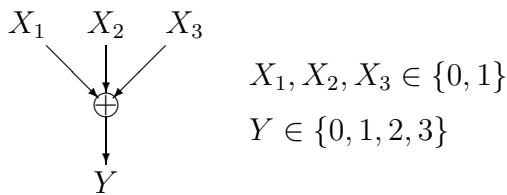

Figure 1: A linear network for which source-channel separation fails [1].

# 1 Introduction

The failure of source-channel separation in networks is often considered to be an impediment in applying information theoretic tools in network settings. The simple multiple access channel of Figure 1 gives one example of how separation can fail [1]. The receiver's channel output is the integer sum of the binary channel inputs of $m \geq 2$ users, yielding a channel output alphabet of size $m + 1$. Since independent, uniformly distributed input signals fail to achieve the maximum mutual information between the transmitted and received signals, direct transmission of dependent source bits over the channel without channel coding sometimes yields higher achievable transmission rates than Slepian-Wolf source coding followed by multiple access channel coding.

While this simple example may at first appear to establish irrefutably the failure of source-channel separation in networks, its simplicity is misleading. In particular, note that the alphabet size of the output is dependent on the number of transmitters. Thus, the network lacks a consistent digital framework. Replacing integer addition with binary addition to give a channel with input and output alphabets of the same cardinality yields a communication system for which separation holds.

In this paper, we argue that source-channel separation is more robust than counterexamples may suggest. We assert, however, that separate source and channel code design does not necessarily simplify the design of communication systems for digital networks. The operations of compression and channel coding are conceptual tools rather than necessary components. While modularity, such as that afforded by the separation theorem, is desirable in the design of components, the decomposition of a problem into modular tasks may increase complexity when the decomposition imposes unnecessary constraints.

In addition to examining traditional questions of source-channel separation, we also investigate a variety of other separation assumptions implicit in common network design techniques. By assuming independent data bits and lossless links, layered approaches to network design endorse a philosophy where source and channel coding are separated from network coding or routing. Through examples, we demonstrate the fragility of this assumed separation. Even in simple digital networks, neither separate source-network coding strategies nor separate channel-network coding techniques guarantee optimal communication performance.

Our network model requires a common finite field alphabet at all nodes but allows noise in the form of erasures or additive noise. We treat two important types of canonical networks: noisy multiple access networks, such as may arise in wireless transmissions, and degraded erasure broadcast networks, such as may arise in wireline broadcasting with packet losses due to congestion. These networks are not only some of the fundamental building blocks of network information theory, but also generally demonstrate the breakdown of separation between source coding and coding over the channel or, rather, network. We establish a simple and economical framework and use that framework to first re-derive classical results for source compression and coding for the single transmitter, single receiver channel. We then use the same framework to prove that linear codes are sufficient and asymptotically optimal for the noisy multiple access and erasure degraded broadcast networks. We also show that optimal code construction for these networks is particularly simple. Our approach may be viewed, in the simplest way, as a generalization of information theoretic results known for single-receiver source codes and for single-transmitter, single-receiver channel codes. From the networking perspective, our results bear the interpretation that separate optimization of compression, channel coding, and routing fails to achieve the optimal performance.

For multiple access networks, we show that source-channel separation is optimal for input-independent noise which may be additive or in the form of erasures. Using this property, we prove the optimality of linear codes for multiple access networks. However, separation may fail to achieve the optimal performance when additive noise is input-dependent. For the additive noise channel over the binary field, we compute the maximum difference between the sum channel capacities when channel encoding is done with complete collaboration between the channel encoders and with no collaboration between the channel encoders . We also obtain an expression for the probability that the two sum capacities differ for a binary additive noise multiple access channel picked randomly from the ensemble of all channels of this class. We present an optimal multiple access systematic channel code construction and provide coding techniques when transmissions are bursty.

Though source-channel separation may not hold in general for broadcast channels, we show that it does hold for the erasure broadcast channel. Thus, for erasure broadcast channels, we show that linear codes are also optimal.

Finally, while the multiple access and broadcast networks considered here are important in their own right, we show that we cannot concatenate them arbitrarily and maintain end-to-end functionality. In effect, there may not be separation of large networks into canonical elements. We argue that this lack of separation may pose a real challenge in communication-network system design.

In section 2, we consider source-channel separation for multiple access networks. We discuss background, establish preliminaries and re-derive classical results for single-transmitter, single-receiver networks in section 3. In sections 4 and 5, we consider multiple access networks and broadcast networks, respectively. We address the issue of decomposability of a network into canonical elements and conclude in section 6.

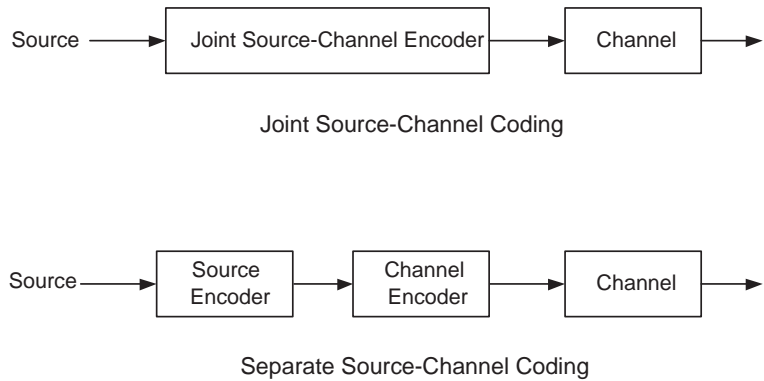

Figure 2: Various single-transmitter, single-receiver schemes.

## 2 Source-channel separation for multiple access networks

It is well known that source-channel separation holds for single-transmitter, single-receiver networks. Thus, the source and channel coding operations can be separated without loss in optimality. The joint and separate source-channel coding schemes for single-transmitter, single-receiver networks are shown in Figure 2.

We now address the topic of separation for source-multiple access channel pairs. Consider binary source pairs $(U_1, U_2)$ and two transmitters transmitting binary symbols to a single receiver whose received alphabet is also binary. We denote the channel inputs, their associated rates and output symbol as $(X_1, X_2)$, $(R_1, R_2)$ and $Y$, respectively.

Let us summarize some known results on multiple access capacity regions. There are three categories of multiple access:

- The most general multiple access is when the channel encoding is done with full collaboration between the channel encoders. Optimal source coding can be performed with or without [22] collaboration between the source encoders. Moreover, there is no loss in optimality in separating the source and channel encoding operations since full collaboration exists between the two channel encoders. We will call this multiple access scheme as "Collaborative Multiple Access" (CMA). The capacity region for this type of multiple access is derived by Liao [7] in his PhD thesis.

- The second type of multiple access is when the source and channel coding at each transmitter is combined into a single operation and there is no collaboration between the joint source-channel encoders at the two transmitters. The encoders directly map the source pairs to channel inputs. We will refer to this multiple access scheme as "Non-collaborative Joint Multiple Access" (NJMA) and the capacity region for this scheme is derived by Cover and El Gamal [2].

- The least general, but most often considered, is multiple access where source and channel coding are performed separately at each transmitter and there is no collaboration between the encoders at the transmitters. We will refer to this multiple access scheme as the "Non-collaborative Separate Multiple Access" (NSMA) and the capacity region for this multiple access scheme is derived by Cover and Wyner [1, 8].

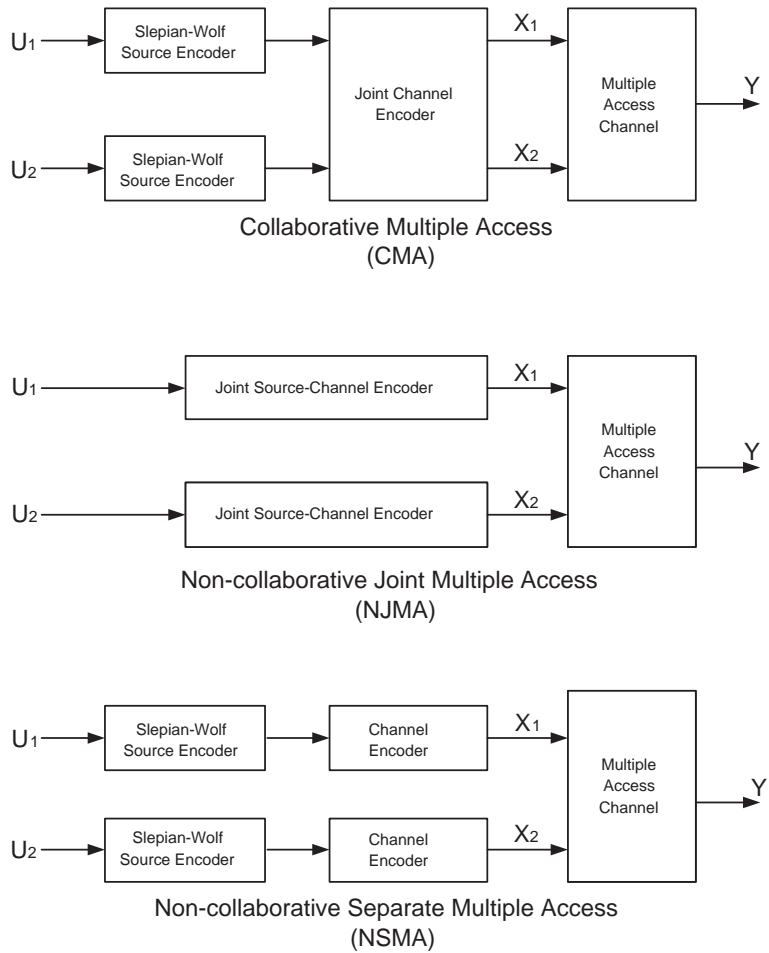

Figure 3: Various multiple access schemes.

The three multiple access capacity schemes are shown in Figure 3. Note that the figure for NSMA is same as Figure 1 in [8].

## 2.1 CMA capacity region

For this type of multiple access, the channel encoders cooperate with each other and can generate any joint input probability distribution, $P_{X_1 X_2}(x_1, x_2)$. For any $P_{X_1 X_2}(x_1, x_2)$, denote the closure of the convex hull of all rate pairs $(R_1, R_2)$ satisfying

$$R_1 < I(X_1; Y | X_2),$$
$$R_2 < I(X_2; Y | X_1),$$
$$R_1 + R_2 < I(X_1, X_2; Y),$$

as $\mathbb{L}[P_{X_1X_2}(x_1,x_2)]$. The CMA capacity region, $\mathbb{R}_{\mathsf{CMA}}$, is the convex hull of the sets $\mathbb{L}[P_{X_1X_2}(x_1,x_2)]$ over all joint input probability distributions, $P_{X_1X_2}(x_1,x_2)$. Denoting the convex hull operation over sets as $\mathcal{CH}(.)$, we have

$$\mathbb{R}_{\mathsf{CMA}} = \mathop{\mathcal{CH}}_{\forall P_{X_1X_2}(x_1,x_2)} \bigg(\mathbb{L}[P_{X_1X_2}(x_1,x_2)]\bigg). \tag{1}$$

As the encoders cooperate in this multiple access scheme, we also refer to the CMA capacity as the "cooperative capacity".

## 2.2 NSMA capacity region

In this multiple access scheme, the channel encoders cannot cooperate and have independent inputs which come, for example, from Slepian-Wolf source encoders. Let $P_{X_1}(x_1)$ and $P_{X_2}(x_2)$ be the distributions on the two independent channel inputs. For any product distribution, $P_{X_1}(x_1)P_{X_2}(x_2)$, denote the closure of the convex hull of all rate pairs $(R_1, R_2)$ satisfying

$$R_1 < I(X_1; Y|X_2),$$
$$R_2 < I(X_2; Y|X_1),$$
$$R_1 + R_2 < I(X_1, X_2; Y),$$

as $\mathbb{C}[P_{X_1}(x_1)P_{X_2}(x_2)]$. The NSMA capacity region, $\mathbb{R}_{\mathsf{NSMA}}$, is the convex hull of the sets $\mathbb{C}[P_{X_1}(x_1)P_{X_2}(x_2)]$ over all product input probability distributions, $P_{X_1}(x_1)P_{X_2}(x_2)$. Hence, we have

$$\mathbb{R}_{\mathsf{NSMA}} = \mathop{\mathcal{CH}}_{\forall (P_{X_1}(x_1), P_{X_2}(x_2))} \bigg(\mathbb{C}[P_{X_1}(x_1)P_{X_2}(x_2)]\bigg). \tag{2}$$

Owing to lack of cooperation, the channel encoders cannot increase the correlation between the inputs which results in the channel inputs being always independent. This makes the NSMA capacity region an improper[1] subset of the CMA capacity region since all joint input distributions cannot be generated. As the encoders in this multiple access scheme do not cooperate, we also refer to the NSMA capacity as the "separate capacity".

## 2.3 NJMA capacity region

In this multiple access scheme, there is a single joint source-channel encoder at each transmitter that maps source symbols to channel inputs. The encoders at the two transmitters do not cooperate. This encoder is more general than the combination of the NSMA source and channel encoders, since it can make use of the dependence between the sources to increase the channel mutual information. As the set of channel input distributions that can be generated is larger than that of the NSMA scheme, the NSMA capacity region is an improper subset of the NJMA capacity region. Also, the NJMA capacity region is an improper subset of the CMA capacity region, since the channel encoders cannot generate all joint input probability distributions, owing to lack of coordination. Only those joint input probability distributions that do not require the correlation between channel inputs to be more than the correlation between the source pairs can be generated. Therefore, the set of joint input distributions that can be generated depends on the source that is being transmitted. We denote the set of joint input distributions that can be generated as $\mathbb{P}^J_{X_1X_2}$.

---

[1] In this paper an improper subset (superset) of a set $\mathcal{A}$ is defined as a set which is smaller (greater) or equal to $\mathcal{A}$.

For any $P_{X_1X_2}(x_1,x_2) \in \mathbb{P}^J_{X_1X_2}$, denote the closure of the convex hull of all rate pairs $(R_1,R_2)$ satisfying

$$R_1 < I(X_1;Y|X_2),$$
$$R_2 < I(X_2;Y|X_1),$$
$$R_1 + R_2 < I(X_1,X_2;Y),$$

as $\mathbb{J}[P_{X_1X_2}(x_1,x_2)]$. The NJMA capacity region, $\mathbb{R}_{\mathsf{NJMA}}$, is the convex hull of the sets $\mathbb{J}[P_{X_1X_2}(x_1,x_2)]$ over all joint input probability distributions in $\mathbb{P}^J_{X_1X_2}$. Hence, we have

$$\mathbb{R}_{\mathsf{NJMA}} = \mathcal{CH}_{\forall P_{X_1X_2}(x_1,x_2) \in \mathbb{P}^J_{X_1X_2}} \left( \mathbb{J}[P_{X_1X_2}(x_1,x_2)] \right). \tag{3}$$

In this paper, we will also refer to this capacity region as the "Joint source-channel capacity region".

## 2.4  A sufficient criterion for separation to hold

Consider the transmission of a binary source pair over a multiple access channel. Even if we allow cooperation, there is no hope of transmitting the source pair over the channel with an arbitrarily small error probability unless the Slepian-Wolf source coding region and the CMA capacity region have a non-zero intersection. The question of source-channel separation therefore applies to those source-channel pairs for which these two regions overlap. Hence, for any channel, while considering source-channel separation, we *always* restrict our attention to source pairs whose Slepian-Wolf region overlaps the CMA capacity region of that channel. If source and channel coding are done separately without coordination between the two transmitters, then only those source pairs for which the Slepian-Wolf region overlaps with the NSMA capacity region can be reliably transmitted over the multiple access channel.

In NJMA, the encoders can make use of the correlation between the sources and hence the NJMA capacity region is an improper superset of the NSMA capacity region. However, since the encoders at the transmitters do not coordinate, this region is an improper subset of the CMA capacity region, in general.

Figure 4 shows the CMA and NSMA capacity regions for a multiple access channel. For ease of illustration, we have considered a multiple-access channel whose capacity regions are pentagons. The regions, in general, may not be pentagons. The Slepian-Wolf regions for three different source pairs are also shown. $ABCDO$ is the NSMA capacity region and $PQRSO$ the CMA capacity region. Source-channel separation holds for all source pairs whose Slepian-Wolf region overlaps $ABCDO$. For source pairs whose Slepian-Wolf region overlaps only $PQRSDCBA$, separation may fail. Separation fails for those source pairs for which the NJMA joint source-channel encoder can increase the capacity region beyond the NSMA capacity region so that it intersects the source pair's Slepian-Wolf source coding region, which allows the source pair to be reliably communicated. The joint source-channel encoders make use of the source statistics to increase correlation between the channel inputs, which may increase the capacity region. More simply, the joint source-channel encoders try to match the source statistics to those needed by the channel to maximize mutual information. However, when the joint source-channel encoders cannot increase the capacity region enough to overlap the source pair's Slepian-Wolf region, reliable communication is impossible. In this case we say that separation holds since joint and separate source-channel codes fail equally. Note that the NJMA capacity region cannot increase beyond the CMA capacity region. Separation fails for the example in [2], since the source pair statistics are perfectly matched to what is required to maximize the channel mutual information. Source pairs whose Slepian-Wolf regions lie outside $PQRSO$ cannot be reliably transmitted over the channel with or without coordination between the transmitters.

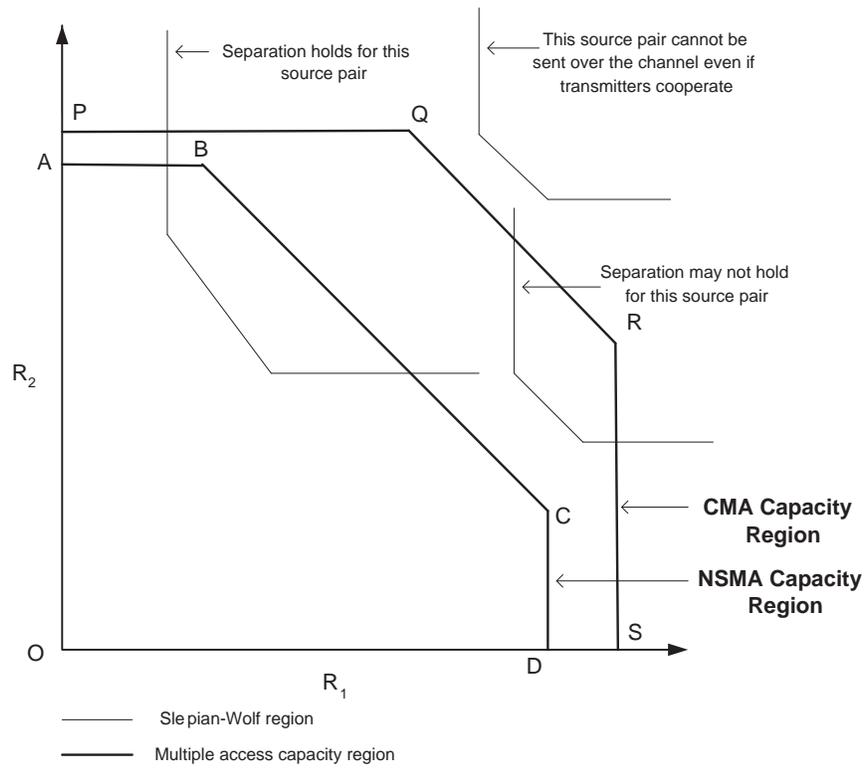

Figure 4: Cooperative and separate multiple access capacity regions.

We now derive a sufficient criterion for separation to hold. Since the NJMA capacity region is an improper subset of the CMA capacity region, a sufficient condition for separation to hold for any source-channel pair is that the NSMA capacity region for the channel is the same as its CMA capacity region. For these channels, increasing correlation between the channel inputs does not increase mutual information. Note that for these channels, the region $PQRSDCBA$ is a null set. We obtain a lemma that states this sufficient criterion for separation to hold for any source-channel pair:

**Lemma 1** *Separation holds for a multiple access source-channel pair if for the channel the following is satisfied*

$$\mathbb{R}_{\mathsf{NSMA}} = \mathbb{R}_{\mathsf{CMA}}.$$

If the sources are independent, joint source-channel coding is equivalent to separate source-channel coding. This yields the following lemma:

**Lemma 2** *Separation holds for a multiple access source-channel pair if the sources are independent.*

## 3 Background, Preliminaries, Single Transmitter-Single Receiver Networks

### 3.1 Background

The use of random linear transformations in coding appears early in the information-theoretic literature. For channel coding, Elias [3] shows that random linear parity check codes, formed by Bernoulli(1/2) choices for the parity check entries in a systematic code's generator matrix, achieve capacity for the binary erasure channel and the binary symmetric channel. Elias also gives a construction for sliding parity check codes requiring fewer random binary digits. MacKay [4] proves that two families of error-correcting codes based on very sparse random parity check matrices – Gallager codes and MacKay-Neal codes (a special case of the former) – when optimally decoded, achieve information rates up to the Shannon limit for channels with symmetric stationary ergodic noise. MacKay also demonstrates empirically, for binary symmetric channels and Gaussian channels, that good decoding performance for these codes can be achieved with a practical sum-product decoding algorithm.

Linear channel coding for network systems has received far less attention. In this work, we consider both multiple access and degraded broadcast channels. Multiple access networks comprise a collection of transmitters sending information to a single receiver. In our network model, the received signal is the sum of the transmitted signals with the possible inclusion of either erasures or additive noise. While this type of additive interference channel has received considerable attention in the literature (see, for example, [6, 7, 9, 10, 11, 12, 13, 14, 16, 17]) the majority of the work to date considers only the case where the incoming data streams interfere additively in the real field. Notable exceptions are the works of Kautz in [15] and Poltyrev and Snyders in [16]. Kautz introduces superimposed codes in [15], for which the received symbol is the boolean sum ("OR") of the channel inputs and in [16], Poltyrev and Snyders treat a modulo-2 multiple access channel without noise. Both works consider the case where a proper subset of the transmitters transmit to the decoder at any given instant. We are unaware of prior work on linear coding for multiple access channels.

In broadcast networks, we consider physically and stochastically degraded channels with both additive noise and erasures. While the degraded broadcast channel is well understood, [18, 19], we are likewise unaware of any prior work on linear broadcast channel codes.

On the source coding side, Ancheta [20] presents universally optimal linear codes for lossless coding of binary sources in point-to-point networks; he also shows that the rate distortion function of a binary, stationary, memoryless source cannot be achieved by any linear transformation over a binary field into a sequence with rate lower than the entropy of the source. The syndrome-source-coding scheme described by Ancheta uses a linear error correcting code for data compression, treating the source sequence as an error pattern whose syndrome forms the compressed data.

In [21], Csiszár generalizes linear source coding techniques to allow linear multiple access source codes that achieve the optimal performance derived by Slepian and Wolf [22]. Csiszár demonstrates the universality of his proposed linear codes[2] and bounds the corresponding error exponents. The linear coding results are generalized for more than two sources and receivers in [39]. These results are generalizable to single or multiple Markov sources.

Addressing the problem of practical encoding and decoding for multiple access source codes, [23, 24, 25, 26, 27] introduce the Distributed Source Coding Using Syndromes (DISCUS) framework, initially looking at sources with strongly structured statistical dependencies. Schonberg et al. [28] note that Csiszár's proof can be used to show that application of LDPC codes in the DISCUS framework approaches the Slepian-Wolf bound for general binary sources; they then demonstrate through simulation that belief propagation decoding works well in practice, with a small performance gap due to the finite block length and choice of parity check matrix. In [40], it is shown that LDPC codes can achieve any point in the Slepian-Wolf region with optimal decoding. Uyematsu proposes a deterministic construction for linear multiple access source codes in [29]; the resulting codes achieve any point in the achievable rate region, with two-step encoding and decoding procedures (similar to concatenated codes for channel coding) of complexity polynomial in the block length.

---

[2]In the given fixed-rate coding regime, a universal code is any code that achieves asymptotically negligible error probability on all sources for which the code's rate falls within the source's achievable rate region.

In other related work, multiple access source code design by randomly choosing among general block codes is considered as an exercise in [30]. Loeliger [31] considers averaging for sets of linear codes with basic symmetry properties and gives a general version of the Varshamov-Gilbert bound and a random coding bound that depend only on the size of the set of error patterns; these results extend corresponding prior results for more specific types of error patterns. Among the applications mentioned are burst error correction, and multiple access systems where each user considers the set of possible interference patterns arising from the activity of other users as well as channel noise.

Zhao and Effros introduce broadcast system source codes in [32, 33]. In a broadcast system source code, a single encoder describes multiple sources to be decoded at a collection of receivers. Sources may include both "common information" intended for more than one receiver and "specific information" intended for only one receiver. In the most general case, they allow a distinct source for every non-empty subset of the set of possible receivers. Design algorithms and performance bounds for lossless broadcast system source codes appear in [32, 33].

Network coding, introduced in [34], is a generalization of routing for transmitting bits through lossless networks. The sufficiency of linear network codes for multicast networks is shown in [35, 36] whereas Koetter and Médard give an algebraic framework in [37]. Reference [38] considers a randomized approach for independent or linearly correlated sources, while [41] and [43] give polynomial-time deterministic and randomized network code constructions for independent sources. Chou et al. [42] demonstrate the practical use of random linear codes over the network topologies of commercial Internet Service Providers.

## 3.2 Preliminaries

Since the focus of our paper is on the relationships between system components and concepts, we give all results in their simplest forms. In particular, we state our results and their corresponding derivations for independent, identically distributed (i.i.d) random processes and focus on binary source and channel alphabets, modified only for the inclusion of the erasure noise model. For simplicity, all code constructions combine random linear encoding with typical set decoding. The definition of the typical set $A_\epsilon^{(n)}$ for a single random sequence $U_1, U_2, \ldots$ drawn i.i.d according to probability mass function (pmf) $p$ is

$$A_\epsilon^{(n)} = \left\{ u^n \in \mathcal{U}^n : H(U) - \epsilon < -\frac{1}{n} \log p(u^n) < H(U) + \epsilon \right\}.$$

Given source alphabet $\mathcal{U}$, $H(U) = -\sum_{u \in \mathcal{U}} p(u) \log p(u)$ is the entropy of the i.i.d random process $U_1, U_2, \ldots$. By the Asymptotic Equipartition Property (AEP),

$$|A_\epsilon^{(n)}| \leq 2^{n(H(U)+\epsilon)}$$

and $\Pr(U^n \in A_\epsilon^{(n)}) \to 1$ as $n \to \infty$. In most cases, we use context to distinguish between typical sets. Thus $U^n \in A_\epsilon^{(n)}$ refers to the typical set for the pmf $p(u)$ of random variable $U$ while $Z^n \in A_\epsilon^{(n)}$ refers to the typical set for the pmf $q(z)$ of random variable $Z$. Focusing on linear encoding and typical set decoding allows us to include the corresponding proofs and illuminates the relationships between them.

While we state and prove our results in their simplest form for readability, we note that all of the results given here generalize widely from the forms that we state explicitly. Some of these generalizations are described below.

- While we focus on the binary alphabet, results generalize to arbitrary finite fields[3]. The requirement that the finite field be the same for all sources, channel codewords, and additive noise processes cannot, however, be relaxed in general. The channel output alphabet is allowed to differ only in the inclusion of erasures. In our model, erasures propagate as erasures when the output of one channel is fed into the input of a subsequent channel.

- We state results for i.i.d source and noise random processes; the results generalize to stationary, ergodic processes for which corresponding typical sets exist.

- We use distribution-dependent typical set decoders; many of the results in this paper can be generalized to achieve universal coding performance and improved error exponents using the maximal entropy decoders of Csiszár [21].

- We ignore decoder complexity issues; good (sub-optimal) decoders with lower complexity can be derived for many of the systems described here using sparse matrix techniques like the the low-density parity-check (LDPC) coding techniques developed by Gallager [44], MacKay [4], Urbanke et al. [5], and others.

- We give results for the smallest generalizable instances of each network type (e.g., two-receiver broadcast channels and three-receiver broadcast system source codes); our results generalize to larger systems.

## 3.3 Single-Transmitter, Single-Receiver Networks

We begin by examining simple forms of some of the prior results described in Section 3.1. In particular, we give simple new proofs for the linear source and channel coding theorems for single-transmitter, single-receiver networks [3, 20, 21]. Our goal is to integrate, in a single simple framework, the earlier known results for linear source and channel coding. These new derivations demonstrate the relationships between these algorithms and random linear network coding techniques. We further provide a linear source coding converse. Finally, we extend the given random design arguments to design linear joint source-channel codes for the single-transmitter, single-receiver network.

### 3.3.1 Linear Source Coding

Given a single-transmitter, single-receiver network, source coding is equivalent to network coding of compressible source sequences. We say that a network code accomplishes optimal source coding on a noise-free network if that code can be used to transmit any source with entropy lower than the network capacity with asymptotically negligible error probability.

---

[3]The results are not restricted to finite fields and hold even when the alphabet possesses a ring structure.

Shannon's achievability result for lossless source coding demonstrates that for $U_1, U_2, \ldots$ drawn i.i.d from a Bernoulli($p$) distribution and any $\epsilon > 0$, there exists a fixed-rate-$(H(U)+\epsilon)$ code for which the probability of decoding error can be made arbitrarily small as the coding dimension $n$ grows without bound. The converse to Shannon's source coding theorem states that asymptotically negligible error probabilities cannot be achieved with rates lower than $H(U)$. We begin by showing that the expected error probability of a randomly chosen, rate-$R$, linear source code approaches zero as $n$ grows without bound for any source $U$ with $H(U) < R$.

**Theorem 1** *Let $U_1, U_2, \ldots, U_n$ be drawn i.i.d according to distribution $p(u)$ on $\mathbb{F}_2$. For any rate, $R > H(U)$, the error probability averaged over the ensemble of random linear source codes tends to 0 as the codeword length tends to $\infty$.*

*Proof:* The fixed-rate, linear encoder is independent of the source distribution. We use distribution-dependent typical set decoders for simplicity. We first describe the source encoder and decoder for a fixed linear code.

Let $A_n$ be an $\lceil nR \rceil \times n$ matrix with coefficients in the binary field $\mathbb{F}_2$. The encoder for the linear source code based on $A_n$ is
$$\alpha_n(u^n) = A_n \mathbf{u},$$
where $u^n = \mathbf{u}^t \in (\mathbb{F}_2)^n$ is an arbitrary source sequence with blocklength $n$. The corresponding decoder is
$$\beta_n(v^{\lceil nR \rceil}) = \begin{cases} u^n & \text{if } u^n \in A_\epsilon^{(n)} \text{ and } A_n \mathbf{u} = \mathbf{v} \text{ and } \not\exists \hat{\mathbf{u}}^n \in A_\epsilon^{(n)} \cap \{\mathbf{u}\}^c \text{ s.t. } A_n \hat{\mathbf{u}} = \mathbf{v} \\ \hat{U}^n & \text{otherwise,} \end{cases}$$
where $v^{\lceil nR \rceil} = \mathbf{v}^t \in (\mathbb{F}_2)^{\lceil nR \rceil}$ and decoding to $\hat{U}^n$ denotes a random decoder output (which yields a decoding error by assumption). The error probability for source code $A_n$ is
$$P_e(A_n) = \Pr(\beta_n(\alpha_n(U^n)) \neq U^n).$$

We design a sequence $\{A_n\}_{n=1}^\infty$ of codes at random and show that if the rate is chosen appropriately, then the expected error probability $E[P_e(A_n)]$ of the randomly chosen code decays to zero as $n$ grows without bound. Using the above encoder and decoder definitions and letting $\mathbf{w}^t \in \mathbb{F}_2^n$ be an arbitrary nonzero vector,

$$\begin{aligned}
E[P_e^{(n)}] &= E[\Pr(\beta_n(\alpha_n(U^n)) \neq U^n)] \\
&= \sum_{u^n \notin A_\epsilon^{(n)}} p(u^n) \Pr(\beta_n(\alpha_n(u^n)) \neq u^n) + \sum_{u^n \in A_\epsilon^{(n)}} p(u^n) \Pr(\beta_n(\alpha_n(u^n)) \neq u^n) \\
&\leq \epsilon_n + \sum_{u^n, \hat{u}^n \in A_\epsilon^{(n)}} p(u^n) \mathbf{1}(\hat{\mathbf{u}} \neq \mathbf{u}) \Pr(A_n \hat{\mathbf{u}} = A_n \mathbf{u}) & (4) \\
&\leq \epsilon_n + \sum_{u^n \in A_\epsilon^{(n)}} p(u^n) 2^{n(H(U)+\epsilon)} \Pr(A_n \mathbf{w} = \mathbf{0}) & (5) \\
&\leq \epsilon_n + 2^{n(H(U)+\epsilon)} 2^{-\lceil nR \rceil} & (6)
\end{aligned}$$

for some $\epsilon_n \to 0$. Equation (4) and the bound on the size of the typical set follow from the AEP. The symmetry represented by the introduction of **w** in (5) and the bound on the corresponding probability in (6) result from the following argument. Let $k$ be the number of ones in an arbitrary $\mathbf{w} \neq \mathbf{0}$. Then each coefficient of vector $A_n \mathbf{w}$ is the sum of $k$ independent Bernoulli(1/2) random variables. Since summing i.i.d Bernoulli(1/2) random variables yields a Bernoulli(1/2) random variable and the rows of $A_n$ are chosen independently, $A_n \mathbf{w}$ is uniformly distributed over its $2^{\lceil nR \rceil}$ possible outcomes.

By (6), $E[P_e^{(n)}] \to 0$ as $n \to \infty$ provided that $\lceil nR \rceil > n(H(U) + \epsilon)$. □

We now present Lemma 3, which provides a form of converse to Theorem 1. While Theorem 1 shows that linear source codes are asymptotically optimal, Lemma 3 shows that any fixed non-trivial linear code yields statistically dependent output symbols. The result of this lemma highlights one difference between the fixed-rate, asymptotically lossless linear codes investigated here and the more typically applied variable-rate, truly lossless source coding schemes like Huffman and arithmetic codes. Variable-rate schemes can achieve lossless performance for any blocklength and precisely achieve the entropy for dyadic distributions.

**Lemma 3** *Given any $n > 1$, let $p_1, \ldots, p_n$ be non-uniform probability mass functions on the mutually independent random variables $U_1, \ldots, U_n$. Defining $\mathbf{V} = (V_1, \ldots, V_k)^t$ and $\mathbf{U} = (U_1, \ldots, U_n)^t$, let*

$$\mathbf{V} = a\mathbf{U}$$

*for an arbitrary $k \times n$ matrix $a$. If $V_1, V_2, \ldots, V_k$ are mutually independent, then matrix $a$ has at most one non-zero element in each column.*

*Proof:* See Appendix 1. □

An immediate consequence of this observation is the following corollary:

**Corollary 1** *Linear source codes cannot achieve the entropy bound for non-uniform sources.*

This corollary follows from the fact that achieving the entropy bound necessarily yields an incompressible data sequence. We address the advantages of fixed-rate codes later in this section by showing how fixed-rate, linear source and channel codes combine naturally to give linear joint source-channel codes.

### 3.3.2 Linear Channel Coding

Just as source coding can be viewed as an extension of network coding to applications with statistically dependent input symbols, channel coding can be viewed as an extension of network coding to unreliable channels. Prior network coding results address the issue of robust communication over unreliable channels by considering strategies for working with non-ergodic link failures [37, 38]. We here investigate ergodic failures. A network code designed for a single-transmitter, single-receiver network with ergodic failures is a channel code for the erasure channel. We say that a network code accomplishes optimal channel coding on the given channel if the network code can be used to transmit any source with rate lower than the noisy channel capacity with asymptotically negligible error probability.

Shannon's channel coding theorem shows that for any channel with capacity $C$, there exists a rate $C - \epsilon$ code, where $\epsilon > 0$, such that the decoding error probability can be made arbitrarily small as we increase the codeword length. The converse states that asymptotically low error probabilities cannot be obtained for rates above the channel capacity. We now show that random linear channel codes achieve the capacity for the binary erasure and additive noise channels.

**Theorem 2** *Consider an erasure channel with input and output alphabets equal to $\mathbb{F}_2$ and $\{0, 1, \mathsf{E}\}$, respectively. The erasure sequence $Z_1, Z_2, \ldots$ is drawn i.i.d according to distribution $q(z)$, where $Z_i = 1$ denotes the erasure event, and $Z_i = 0$ designates a successful transmission. The channel noise is independent of the channel input by assumption. If the transmission rate is less than the channel capacity, i.e., $R < 1 - q(1)$, then the error probability averaged over the ensemble of random linear channel codes tends to 0 as the codeword length tends to $\infty$.*

*Proof:* See Appendix 1. □

For the binary additive noise channel model, the noise may be viewed either as true noise, or as the signal of another user that has been combined with the desired signal at some node of a network. The second interpretation is only useful when the interfering signal is not i.i.d uniform; we treat interference channels in detail in Section 4. We begin with the channel coding theorem for the additive noise channel.

**Theorem 3** *Consider an additive noise channel with input, output, and noise alphabets all equal to the binary field $\mathbb{F}_2$. Let noise $Z_1, Z_2, \ldots$ be drawn i.i.d according to distribution $q(z)$. The channel noise is independent of the channel input. If the transmission rate $R$ is less than the channel capacity, i.e., $R < 1 - H(Z)$, then the error probability averaged over the ensemble of random linear channel codes tends to 0 as the codeword length tends to $\infty$.*

*Proof:* Let $A_n$ be an $\lceil n(1-R) \rceil \times n$ matrix with coefficients in $\mathbb{F}_2$. For channel coding, $A_n$ plays the traditional role of the parity check matrix. Following Csiszár [21], however, we interpret $A_n$ as a source code on the noise. For any matrix $A_n$, we can design an $n \times \lfloor nR \rfloor$ matrix $B_n$ such that $B_n$ has full rank and $A_n B_n = \mathbf{0}$. Matrix $B_n$ plays the role of the generator matrix for the desired channel code. We design $B_n$ to have full rank so that each length-$\lfloor nR \rfloor$ input message maps to a distinct channel codeword. We force $A_n B_n = \mathbf{0}$ so that each codeword is in the null space of $A_n$, making possible separation of the encoded message from the additive noise.

More precisely, the channel encoder is defined by

$$\gamma(v^{n-k}) = B_n \mathbf{v}.$$

The channel output for a random channel input $B_n \mathbf{V}$ is

$$\mathbf{Y} = B_n \mathbf{V} + \mathbf{Z}.$$

In decoding the channel output, the receiver first multiplies $\mathbf{Y}$ by $A_n$ to give

$$A_n \mathbf{Y} = A_n(B_n \mathbf{V} + \mathbf{Z}) = A_n \mathbf{Z}.$$

The result of this multiplication is a source coded description of the error signal **Z**. Thus the decoding procedure involves applying source decoder $\beta_n$ to $A_n\mathbf{Y}$. The error is decoded correctly with high probability. The receiver then subtracts the error estimate from the received **Y** to yield, with high-probability, $B_n\mathbf{V}$. Since $B_n$ has full rank, the receiver can recover **V** perfectly from $B_n\mathbf{V}$. Thus the channel code's error probability equals the error probability for the corresponding source code on the error signal $Z^n$. Given this insight, the channel coding theorem is an immediate extension of the source coding theorem.

As in the proof of Theorem 1, we choose a sequence $\{A_n\}_{n=1}^{\infty}$ of matrices at random. This is our source code for the noise. For each $A_n$, we design an $n \times (n-k)$ matrix $B_n$ such that $B_n$ has full rank and $A_n B_n = \mathbf{0}_{k \times (n-k)}$. By the argument given above, the error probability for the given channel code equals the error probability for the corresponding source code on the error signal $Z^n$. By Theorem 1, the expected value of this error probability goes to zero as $n$ grows without bound for all $\lceil n(1-R) \rceil > nH(Z)$, giving an asymptotically negligible error probability for any $R < 1 - H(Z)$. □

### 3.3.3 Linear Joint Source-Channel Coding

From the coding theorem for single transmitter-single receiver channels, any binary source $U$ with entropy $H(U)$ can be transmitted over a channel with capacity $C$ with arbitrarily low probability of decoding error as long as $H(U) < C$. Moreover, if $H(U) > C$, the probability of error is bounded away from zero, and it is not possible to send the source process reliably over the channel. We now show that random linear joint source-channel codes achieve capacity for the binary erasure and additive noise channels.

Since source-channel separation holds for the single-transmitter, single-receiver network, concatenating optimal linear source and channel codes yields an optimal linear joint source-channel code. As an alternative to this approach, where we design separate random linear source and channel codes and concatenate them together, we can design a joint source-channel code at random and decode in a single typical set decoding argument. While we stick with the traditional name of joint source-channel coding, we note that the code does not perform the separate functions of source and channel coding jointly. Instead, the code maps source sequences to channel inputs in a manner that allows robust communication without any explicit or implicit compression or addition of channel coding redundancy. We present Theorems 4 and 5, which show that random linear joint source-channel codes are optimal for sending i.i.d Bernoulli sources over the binary erasure channel and binary additive noise channel, respectively.

**Theorem 4** *Consider the random source $U_1, U_2, \ldots$ drawn i.i.d according to distribution $p(u)$, and let $Z_1, Z_2, \ldots$ be the channel's random erasures, where $Z_1, Z_2, \ldots$ are drawn i.i.d according to distribution $q(z)$ and are independent of the source. (Again $Z_i = 1$ denotes an erasure event.) Assume that the source and channel input alphabets are equal to the binary field $\mathbb{F}_2$. If $H(U) < 1 - q(1)$, then the error probability averaged over the ensemble of random linear codes tends to 0 as the codeword length tends to $\infty$.*

*Proof:* See Appendix 1. □

**Theorem 5** *Consider the random source $U_1, U_2, \ldots$ drawn i.i.d according to distribution $p(u)$, and let $Z_1, Z_2, \ldots$ be the channel's random additive noise, where $Z_1, Z_2, \ldots$ are drawn i.i.d according to distribution $q(z)$ and are independent of the source. Assume that the source, channel input, channel output, and noise alphabets are all equal to the binary field $\mathbb{F}_2$. If $H(U) < 1 - H(Z)$, then the error probability averaged over the ensemble of random linear codes tends to 0 as the codeword length tends to $\infty$.*

*Proof:* See Appendix 1. □

## 4 Multiple Access Networks

The techniques applied in the previous section for single-transmitter, single-receiver systems can also be applied to the design of linear source and channel codes for multiple access networks. For these networks, we show that source-channel separation holds when noise is independent of inputs and prove the optimality of linear joint source-channel codes.

### 4.1 Linear Source Coding

In [22], Slepian and Wolf derive the rate region for multiple access source codes. Csiszár generalizes linear source coding techniques in [21] to show that linear multiple access source codes achieve all points in the rate region with arbitrarily small probability of error. We begin with a simple and short re-derivation of the linear multiple access source codes first studied by Csiszár. Our goal is to integrate the known results on source coding into our framework.

**Theorem 6** *Consider source sequence $(U_{1,1}, U_{2,1}), (U_{1,2}, U_{2,2}), \ldots$ drawn i.i.d according to distribution $p(u_1, u_2)$ on $(\mathbb{F}_2)^2$. Then for any rates*

$$\begin{aligned} R_1 &> H(U_1|U_2) \\ R_2 &> H(U_2|U_1) \\ R_1 + R_2 &> H(U_1, U_2), \end{aligned}$$

*the error probability averaged over the ensemble of random linear multiple-access source codes tends to 0 as the codeword lengths tend to $\infty$.*

*Proof:* Given $\lceil nR_1 \rceil \times n$ matrix $A_{1,n}$ and $\lceil nR_2 \rceil \times n$ matrix $A_{2,n}$, we associate with $(A_{1,n}, A_{2,n})$ a blocklength-$n$, two-transmitter, linear multiple access source code as follows. For any $u_1^n = \mathbf{u}_1^t \in (\mathbb{F}_2)^n$ and $u_2^n = \mathbf{u}_2^t \in (\mathbb{F}_2)^n$, encoders 1 and 2 are defined by

$$\begin{aligned} \alpha_{1,n}(u_1^n) &= A_{1,n}\mathbf{u}_1 \\ \alpha_{2,n}(u_2^n) &= A_{2,n}\mathbf{u}_2. \end{aligned}$$

For any $v_1^{\lceil nR_1 \rceil} = \mathbf{v}_1^t \in (\mathbb{F}_2)^{\lceil nR_1 \rceil}$ and $v_2^{\lceil nR_2 \rceil} = \mathbf{v}_2^t \in (\mathbb{F}_2)^{\lceil nR_2 \rceil}$, the decoder is defined by

$$\beta_n(v_1^{\lceil nR_1 \rceil}, v_2^{\lceil nR_2 \rceil}) = \begin{cases} (u_1^n, u_2^n) & \text{if } (u_1^n, u_2^n) \in A_\epsilon^{(n)} \text{ and } (A_{1,n}\mathbf{u}_1, A_{2,n}\mathbf{u}_2) = (\mathbf{v}_1, \mathbf{v}_2) \text{ and} \\ & \nexists (\hat{\mathbf{u}}_1, \hat{\mathbf{u}}_2) \in A_\epsilon^{(n)} \cap \{(\mathbf{u}_1, \mathbf{u}_2)\}^c \text{ s.t.} \\ & (A_{1,n}\hat{\mathbf{u}}_1, A_{2,n}\hat{\mathbf{u}}_2) = (\mathbf{v}_1, \mathbf{v}_2) \\ (\hat{U}_1^n, \hat{U}_2^n) & \text{otherwise.} \end{cases}$$

Again, decoding to $(\hat{U}_1^n, \hat{U}_2^n)$ denotes an error event.

An error occurs if either or both of the source sequences is decoded in error. Thus, following an argument very similar to those seen previously,

$$\begin{aligned}
&E[P_e(A_{1,n}, A_{2,n})] \\
&= E[\Pr(\beta_n(\alpha_{1,n}(U_1^n), \alpha_{2,n}(U_2^n))) \neq (U_1^n, U_2^n) \wedge (U_1^n, U_2^n) \notin A_\epsilon^{(n)}] \\
&\quad + E[\Pr(\beta_n(\alpha_{1,n}(U_1^n), \alpha_{2,n}(U_2^n))) \neq (U_1^n, U_2^n) \wedge (U_1^n, U_2^n) \in A_\epsilon^{(n)}] \\
&\leq \epsilon_n + \sum_{(u_1^n, u_2^n) \in A_\epsilon^{(n)}} p(u_1^n, u_2^n) \sum_{\hat{u}_2^n : (u_1^n, \hat{u}_2^n) \in A_\epsilon^{(n)}} 1(\hat{u}_2^n \neq u_2^n) \Pr(A_{2,n}(\mathbf{u}_2 - \hat{\mathbf{u}}_2) = \mathbf{0}) \\
&\quad + \sum_{(u_1^n, u_2^n) \in A_\epsilon^{(n)}} p(u_1^n, u_2^n) \sum_{\hat{u}_1^n : (\hat{u}_1^n, u_2^n) \in A_\epsilon^{(n)}} 1(\hat{u}_1^n \neq u_1^n) \Pr(A_{1,n}(\mathbf{u}_1 - \hat{\mathbf{u}}_1) = \mathbf{0}) \\
&\quad + \sum_{(u_1^n, u_2^n), (\hat{u}_1^n, \hat{u}_2^n) \in A_\epsilon^{(n)}} p(u_1^n, u_2^n) 1(\hat{u}_1^n \neq u_1^n) 1(\hat{u}_2^n \neq u_2^n) \\
&\qquad \cdot \Pr((A_{1,n}(\mathbf{u}_1 - \hat{\mathbf{u}}_1), A_{2,n}(\mathbf{u}_2 - \hat{\mathbf{u}}_2)) = (\mathbf{0}, \mathbf{0})) \\
&\leq \epsilon_n + 2^{n(H(U_2|U_1)+2\epsilon)} \Pr(A_{2,n}\mathbf{w} = \mathbf{0}) + 2^{n(H(U_1|U_2)+2\epsilon)} \Pr(A_{1,n}\mathbf{w} = \mathbf{0}) \\
&\quad + 2^{n(H(U_1,U_2)+\epsilon)} \Pr(A_{1,n}\mathbf{w}_1 = \mathbf{0} \wedge A_{2,n}\mathbf{w}_2 = \mathbf{0}) \\
&= \epsilon_n + 2^{-(\lceil nR_2 \rceil - n(H(U_2|U_1)+2\epsilon))} + 2^{-(\lceil nR_1 \rceil - n(H(U_1|U_2)+2\epsilon))} \\
&\quad + 2^{-(\lceil nR_1 \rceil + \lceil nR_2 \rceil - n(H(U_1,U_2)+\epsilon))}
\end{aligned}$$

for arbitrary, non-zero $\mathbf{w}^t, \mathbf{w}_1^t, \mathbf{w}_2^t \in \mathbb{F}_2^n$ and some $\epsilon_n \to 0$. Thus for all $(\lceil nR_1 \rceil, \lceil nR_2 \rceil)$ satisfying $\lceil nR_1 \rceil > n(H(U_1|U_2) + 2\epsilon)$, $\lceil nR_2 \rceil > n(H(U_2|U_1) + 2\epsilon)$, and $\lceil nR_1 \rceil + \lceil nR_2 \rceil > n(H(U_1, U_2) + \epsilon)$, $E[P_e(A_{1,n}, A_{2,n})] \to 0$ as $n$ grows without bound. $\square$

This theorem proves the optimality of linear multiple access source codes and re-establishes the results of Csiszár [21].

## 4.2 Linear Channel Coding

Application of linear channel coding techniques to achieve linear multiple access channel codes is more straightforward than the corresponding source coding result. In particular, we consider the two additive multiple access channels shown in Figure 5. The first is the additive multiple access channel with erasures and the second is the additive multiple access channel with additive noise. The additive channel with interference only (no channel noise) can be viewed as a special case of either of the noisy models where errors or erasures occur with probability zero. Let $X_1^n$ and $X_2^n$ denote the random channel inputs and use $Y^n$ to denote the corresponding random channel output. $Y^n$ equals $X_1^n + X_2^n$ corrupted by erasures in the

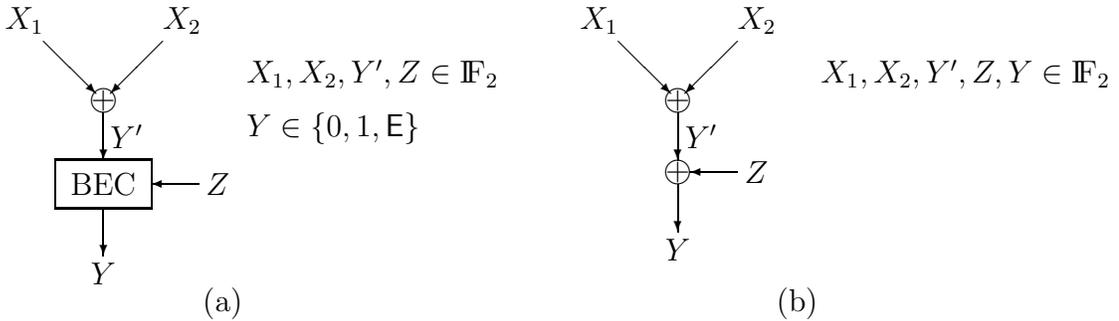

Figure 5: Binary additive multiple access channels with (a) erasures and (b) additive noise. In both cases, $Z_1, Z_2, \ldots$ are i.i.d and independent of the channel inputs.

erasure channel model; we denote the probability of an erasure as $q(1)$. For the additive noise channel model, $Y^n$ equals $X_1^n + X_2^n + Z^n$, where $Z^n$ is the i.i.d additive binary noise. Both examples use addition over the binary field; and noise is independent of the channel inputs. We begin by deriving the NSMA capacity regions of the multiple access channel with erasures and the multiple access channel with additive noise. The following lemma describes these regions:

**Lemma 4** *The NSMA capacity region for both the additive multiple access channel with erasures and the additive multiple access channel with additive noise equal the rate region achieved by time-sharing between the points $(C, 0)$ and $(0, C)$, where $C = 1 - q(1)$ for the erasure model and $C = 1 - H(Z)$ for the additive noise model.*

*Proof:* See Appendix 2. □

Since time-sharing between two linear codes can itself be described as a linear code, the time-sharing solution demonstrates not only that the end points are achievable by linear codes but also that all points in the set of achievable rates are achievable by linear multiple access channel codes. Therefore, we have Theorems 7 and 8 that random linear multiple access channel codes achieve the NSMA capacity for the binary multiple access channel with erasures and the binary multiple access channel with additive noise, respectively.

**Theorem 7** *Consider a multiple access channel with input alphabets $\mathcal{X}_1 = \mathcal{X}_2 = \mathbb{F}_2$ and output alphabet $\mathcal{Y} = \{0, 1, \mathsf{E}\}$. If the channel inputs at time $i$ are $X_{1,i}$ and $X_{2,i}$, then the channel output at time $i$ is the binary sum $X_{1,i} + X_{2,i}$ with probability $q(0)$ and $E$ with probability $q(1)$. Erasures are i.i.d and independent of the channel inputs. The error probability averaged over the ensemble of rate-$(R_1, R_2)$ random linear multiple access channel codes tends to 0 as the codeword lengths tend to $\infty$, if $R_1 + R_2 < 1 - q(1)$.*

*Proof:* See Appendix 2. □

**Theorem 8** *Consider a multiple access channel with input-independent, additive noise. Suppose that the input alphabets, output alphabet, and noise alphabet are all equal to the binary field $\mathbb{F}_2$. Let noise $Z_1, Z_2, \ldots$ be drawn i.i.d according to distribution $q(z)$. If the channel* inputs at time $i$ are $X_{1,i}$ and $X_{2,i}$, then the channel output at time $i$ is $Y_i = X_{1,i} + X_{2,i} + Z_i$. The error probability averaged over the ensemble of rate-$(R_1, R_2)$ random linear multiple access channel codes tends to 0 as the codeword lengths tend to $\infty$, if $R_1 + R_2 < 1 - H(Z)$.

*Proof:* See Appendix 2. □

## 4.3 Linear Joint Source-Channel Coding

We start by showing that source-channel separation holds for binary sources and binary erasure or additive noise multiple access channels where the erasure or additive noise is independent of channel inputs. This result is embodied in the following theorem that uses Lemma 1 from section 2 in its proof:

**Theorem 9** *For any pair of binary sources and any binary erasure or additive noise multiple access channel where the erasure or additive noise is independent of the channel inputs, separation holds.*[4]

*Proof:* See Appendix 2. □

This leads to Theorems 10 and 11, which apply to the binary multiple access channel with erasures and additive noise, respectively:

**Theorem 10** *Consider a multiple access channel with input alphabets $\mathcal{X}_1 = \mathcal{X}_2 = \mathbb{F}_2$ and output alphabet $\mathcal{Y} = \{0, 1, \mathsf{E}\}$. If the channel inputs at time $i$ are $X_{1,i}$ and $X_{2,i}$, then the channel output at time $i$ is the binary sum $X_{1,i} + X_{2,i}$ with probability $q(0)$ and $\mathsf{E}$ with probability $q(1)$; the erasure events are i.i.d and independent of the channel inputs. If source pair $(U_{1,1}, U_{2,1}), (U_{1,2}, U_{2,2}), \ldots$ is drawn i.i.d according to distribution $p(u_1, u_2)$ with $H(U_1, U_2) < 1 - q(1)$, then there exists a sequence of joint source-channel codes with probability of error $P_e^{(n)} \to 0$. Conversely, if $H(U_1, U_2) > 1 - q(1)$, then the probability of error for any communication system is bounded away from zero.*

*Proof:* See Appendix 2. □

**Theorem 11** *Consider a multiple access channel with input-independent, additive noise. Suppose that the input alphabets, output alphabet, and noise alphabet are all equal to the binary field $\mathbb{F}_2$. Let noise $Z_1, Z_2, \ldots$ be drawn i.i.d according to distribution $q(z)$. If source pair $(U_{1,1}, U_{2,1}), (U_{1,2}, U_{2,2}), \ldots$ is drawn i.i.d according to distribution $p(u_1, u_2)$ with $H(U_1, U_2) < 1 - H(Z)$, then there exists a sequence of joint source-channel codes with probability of error $P_e^{(n)} \to 0$. Conversely, if $H(U_1, U_2) > 1 - H(Z)$, then the probability of error is bounded away from zero.*

---

[4]In order to maximize the mutual information between the inputs and output of a multiple access channel with input-independent noise, we need to maximize the entropy of the channel output. Binary addition (XOR) of two independent binary random variables corresponds to circular convolution of their probability mass functions (pmfs). If one of the pmfs is uniform, the binary sum has a uniform distribution which leads to its entropy being maximized. Thus, if the channel inputs are uniform, they maximize the entropy of the channel output for an additive multiple access channel operating over the binary field. It is this property of circular convolution that gives rise to source-channel separation in multiple access networks operating over finite fields.

*Proof:* See Appendix 2.                                                                      □

Since source-channel separation holds for the multiple access channel with input-independent erasures and additive noise, we can combine the optimal linear source and channel codes into a single joint source-channel code to achieve optimal performance.

We now show that instead of concatenating the linear source and channel codes, we may use a random linear code that maps source sequences directly to channel inputs in a manner that allows optimal performance for the binary multiple access channel with input-independent erasures or additive noise. The two linear joint source-channel encoders do not cooperate with each other.

Theorems 12 and 13 show that random linear joint source-channel codes achieve performance equivalent to that given in Theorems 10 and 11, respectively, and are thus optimal for the binary multiple access channel with input-independent erasures or additive noise.

**Theorem 12** *Consider the random source $(U_{1,1}, U_{2,1}), (U_{1,2}, U_{2,2}), \ldots$ drawn i.i.d according to distribution $p(u_1, u_2)$, and let $Z_1, Z_2, \ldots$ be the channel's random erasures, where $Z_1, Z_2, \ldots$ are drawn i.i.d according to distribution $q(z)$, all $Z_i$ are independent of the source, and $Z_i = 1$ denotes an erasure in channel use $i$. Assume that the source and channel input alphabets are equal to the binary field $\mathbb{F}_2$. If $H(U_1, U_2) < 1 - q(1)$, then the error probability averaged over the ensemble of random linear joint source-channel codes tends to 0 as the codeword lengths tend to $\infty$.*

*Proof:* See Appendix 2.                                                                      □

**Theorem 13** *Consider the random source $(U_{1,1}, U_{2,1}), (U_{1,2}, U_{2,2}), \ldots$ drawn i.i.d according to distribution $p(u_1, u_2)$, and let $Z_1, Z_2, \ldots$ be the channel's random additive noise, where $Z_1, Z_2, \ldots$ are drawn i.i.d according to distribution $q(z)$, and $Z_i$ are independent of the source. Assume that the source, channel input, channel output, and noise alphabets are all equal to the binary field $\mathbb{F}_2$. If $H(U_1, U_2) < 1 - H(Z)$, then the error probability averaged over the ensemble of random linear joint source-channel codes tends to 0 as the codeword lengths tend to $\infty$.*

*Proof:* See Appendix 2.                                                                      □

## 4.4  Multiple access networks with input-dependent additive noise

We have seen that for channels where the NSMA capacity region is equal to the CMA capacity region, separation holds for all source pairs. The binary multiple access erasure and additive noise channels with input-independent noise are examples. However, when the two regions are not the same, separation may not hold for all source pairs.

We analyze the binary additive noise multiple access channel and show that the CMA and NSMA capacity regions may not be the same when the noise is allowed to depend on the channel inputs. Define the CMA and NSMA sum capacities, $R_{sum}^{CMA}$ and $R_{sum}^{NSMA}$, respectively,

as

$$\begin{aligned} R_{sum}^{CMA} &\triangleq \max_{\forall P_{X_1 X_2}(x_1, x_2)} I(X_1, X_2; Y), \\ R_{sum}^{NSMA} &\triangleq \max_{\forall (P_{X_1}(x_1), P_{X_2}(x_2))} I(X_1, X_2; Y). \end{aligned}$$

We compute the maximum loss in sum capacity, $(R_{sum}^{CMA} - R_{sum}^{NSMA})$, over the ensemble of all binary additive noise channels. We also obtain the expression for the probability that the two sum capacities are unequal for a channel chosen randomly from the ensemble. Our results are embodied in the following theorems.

**Theorem 14** *For noisy multiple access binary additive noise channels, the maximum difference between the CMA and NSMA sum capacities is $\frac{1}{2}$ bit per channel use.*

*Proof:* See Appendix 3. □

**Theorem 15** *The probability that the CMA and NSMA sum capacities are unequal for a channel chosen randomly from the ensemble of equally likely channels is $\frac{1}{3}$.*

*Proof:* See Appendix 3. □

## 4.5 Systematic channel code constructions for multiple access networks

While the sum rate of a multiple access channel code measures the average number of bits per channel use successfully communicated by all transmitters to the channel's single receiver, it fails to address the question of what fraction of time each transmitter remains silent (in order to avoid causing interference with other transmitters). We next investigate this question for the noise-free binary multiple access channel.

Define the "*code rate*" as the ratio of the number of information bits recovered at the receiver, to the total number of bits sent by the transmitter, in one time slot. Thus, the code rate is a dimensionless quantity with maximal value 1 that represents the overhead (in the form of redundancy) required for reliable communication. (Code rate is maximized when overhead is minimized.) We assume that when transmitters do not transmit, their channel input is 0. The code rate is 1 when there is no multiple access interference and no noise. The noise free multiple access network then becomes equivalent to a point-to-point channel without noise. We give a systematic linear code construction and prove that it achieves maximal code rate and capacity over all codes. We then prove that all codes that achieve the maximal code rate also achieve capacity. We also look at the bursty case when transmitters transmit according to a Bernoulli random process and propose coding techniques to maximize code rate. For such bursty channels, we show that when the information codewords at the input to the channel encoders have the same size, maximal expected code rate is achieved by adding redundancy at the transmitter with a higher probability of transmission and not adding any redundancy at the transmitter with a lower probability of transmission.

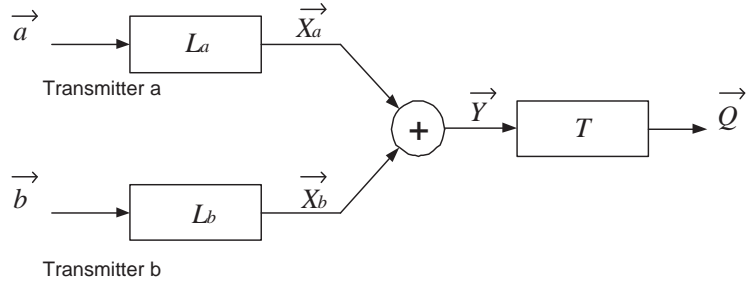

Figure 6: Single Slot Model for the Noise-Free Multiple Access Channel.

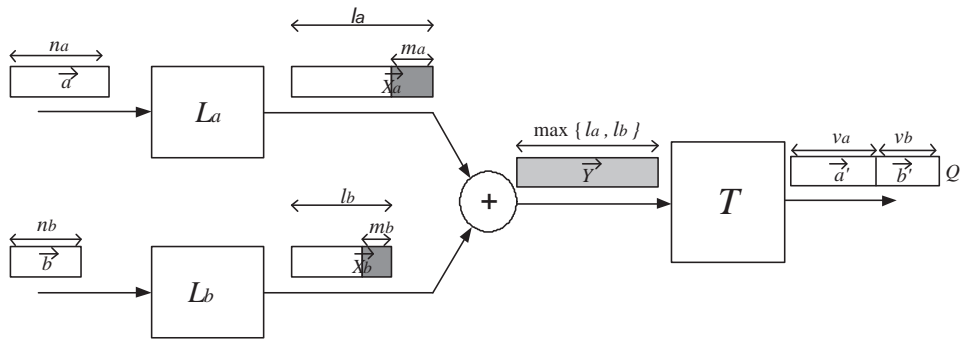

Figure 7: Pictorial representation of communication scheme.

### 4.5.1 Single Slot Model

Since separation holds in binary multiple access additive noise networks when noise is independent of the inputs, it also holds for noise-free binary multiple access channels ($Y = X_a + X_b$ with $X_a, X_b, Y \in \mathbb{F}_2$) where interference from other users limits capacity.

We consider a discrete, time-slotted channel where transmissions start at the beginning of the slot and occur over the length of a slot. The transmitters do not coordinate their transmissions and hence, *time sharing is not possible*. Figure 6 shows a single slot model of the noise-free multiple access channel. Information bits coming out of the source coders in one slot duration are represented as $\vec{a}$ and $\vec{b}$, respectively. Vectors $\vec{a}$ and $\vec{b}$ have sizes $n_a$ and $n_b$ respectively, and have i.i.d and uniformly distributed entries. The elements of $\vec{a}$ and $\vec{b}$ are in $\mathbb{F}_2$. For this model, all operations, matrices and vectors are in the binary field. We refer to $\vec{a}$ and $\vec{b}$ as transmit vectors and assume without loss of generality that $n_a \geq n_b$. Let $m_a$ and $m_b$ be the redundant bits added to vectors $\vec{a}$ and $\vec{b}$, respectively, by the systematic channel code. We use $l_a$ and $l_b$ to denote the lengths of the vectors obtained by channel coding on $\vec{a}$ and $\vec{b}$ respectively, giving $l_a = n_a + m_a$ and $l_b = n_b + m_b$. In general, $l_a \neq l_b$, so both transmitters may not transmit for the entire slot duration. However, at least one transmitter transmits for the whole slot duration. Therefore, the slot length is given by

$$S = \max(l_a, l_b) \text{ bits.} \tag{7}$$

Let $L_a$ be an $(n_a + n_b) \times n_a$ matrix and $L_b$ be an $(n_a + n_b) \times n_b$ matrix. These are the generator matrices for the channel codes at a and b, respectively. Only the first $l_a$ rows of $L_a$ and the first $l_b$ rows of $L_b$ are non-zero. $\vec{X}_a$ and $\vec{X}_b$ are the codewords that are sent over the channel, and they interfere additively over the binary field. At the decoder, a matrix $T$ of dimension $(v_a + v_b) \times (n_a + n_b)$ decodes the received vector to generate a subset of $\vec{a}$ and $\vec{b}$. Let $\vec{Q}$ be the decoded output containing $v_a$ ($v_a \in \{1, \ldots, n_a\}$) bits of $\vec{a}$ denoted by $\vec{a'}$ and $v_b$ ($v_b \in \{1, \ldots, n_b\}$) bits of $\vec{b}$ denoted by $\vec{b'}$. We have the following relations:

$$\vec{X}_a = L_a \vec{a} \qquad \vec{X}_b = L_b \vec{b} \qquad \vec{Y} = L_a \vec{a} + L_b \vec{b},$$
$$\vec{Q} = T\vec{Y} = (TL_a)\vec{a} + (TL_b)\vec{b}.$$

Figure 7 describes the communication scheme pictorially.

### 4.5.2 Maximal Code Rate

In this section, we derive the maximal code rate of the noise-free binary multiple access channel. For the given interference channel, the NSMA capacity region is the set of all rate pairs, $(R_a, R_b)$, satisfying

$$R_a \leq 1, \tag{8}$$
$$R_b \leq 1, \tag{9}$$
$$R_{sum}^{NSMA} = R_a + R_b \leq 1, \tag{10}$$

where the rates are in bits per channel use.

We now find the maximal code rate for this channel. Transmitters a and b transmit $n_a$ and $n_b$ information bits per slot respectively and the codewords transmitted have a length of $l_a$ and $l_b$ bits respectively. The transmission rates are

$$R_a = \frac{n_a}{S},$$
$$R_b = \frac{n_b}{S},$$
$$R_{sum}^{NSMA} = \frac{n_a + n_b}{S}. \tag{11}$$

We assume transmission at a rate within the capacity region, giving

$$n_a + n_b \leq S, \tag{12}$$

by (10, 11). The code rate is a dimensionless quantity given by

$$C_{rate} = \frac{n_a + n_b}{l_a + l_b} \tag{13}$$
$$= \frac{n_a + n_b}{\min(l_a, l_b) + \max(l_a, l_b)} \tag{14}$$
$$= \frac{n_a + n_b}{\min(l_a, l_b) + S} \tag{15}$$
$$\leq \frac{n_a + n_b}{n_b + n_a + n_b}. \tag{16}$$

Equation (15) is due to (7). Expression (16) follows from (12) and the fact that $n_b \leq n_a$ implies $n_b \leq \min(l_a, l_b)$. Thus, we have

$$C_{rate} \leq \frac{n_a + n_b}{n_a + 2n_b}. \tag{17}$$

### 4.5.3 Code Construction

We now construct systematic codes that achieve the capacity and maximal code rate for the binary noise free multiple access channel; we call these codes *optimal codes*. The code rate, transmission rates $R_a$, $R_b$, and sum rate $R_{sum}^{NSMA}$ for our model are given by

$$C_{rate} = \frac{v_a + v_b}{l_a + l_b} \qquad R_a = \frac{v_a}{S},$$
$$R_b = \frac{v_b}{S} \qquad R_{sum}^{NSMA} = \frac{v_a + v_b}{S}.$$

Note that there may be many optimal codes. In Appendix 4, subsection A.4.1, we describe one such construction and prove its optimality. We also establish Theorem 16, which shows that maximal code rate achieving codes are capacity achieving.

**Theorem 16** *For a noise-free binary multiple access channel, codes achieve the maximal code rate if and only if they are capacity achieving and no redundancy is added to the smaller transmit vector.*

*Proof:* See Appendix 4, subsection A.4.2. □

### 4.5.4 The case when transmitters are bursty

In our discussion of systematic code constructions, we assume that each transmitter has a codeword to transmit in each slot. We now consider the case when the channel encoders may not always have an input information codeword to encode.
We assume that each transmitter transmits in a slot according to a Bernoulli process. The lower the probability of transmission, the burstier the transmitter. We would like to know what coding techniques to use in order to obtain the maximal code rate over a large number of transmissions. Intuitively, it can be expected that bursty transmissions will reduce multiple access interference and increase the code rate. Moreover, we should be able to obtain a code rate of 1 (the code rate of a point-to-point noise-free channel) in the limit that one transmitter stops transmitting. In this section, we illustrate coding techniques for bursty multiple access and also show that the limits that we expect actually hold. The result is embodied in the following theorem:

**Theorem 17** *When the information codewords at the input to the channel encoders have the same size, maximal expected code rate is achieved over the noise-free binary multiple access channel by adding redundancy at the less bursty transmitter and not adding any redundancy at the more bursty transmitter.*

*Proof:* See Appendix 4, subsection A.4.2. □

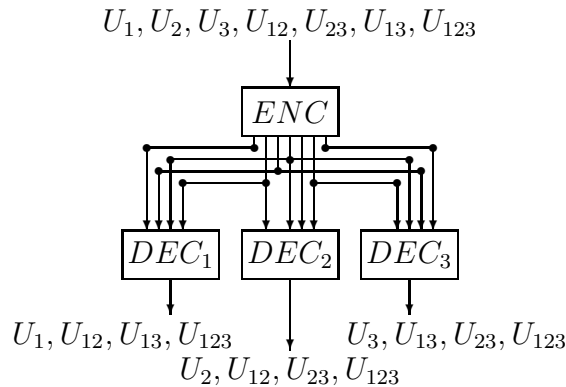

Figure 8: A broadcast system source code with three receivers.

# 5  Broadcast Networks

The next simple model under consideration is the broadcast system, where one transmitter sends information to a collection of receivers. We consider the binary erasure channel and show that linear joint source-channel codes are optimal.

## 5.1  Linear Source Codes

A broadcast system source code comprises a single encoder and a collection of decoders. Since the case with two receivers has special structure absent from general broadcast system source codes [32, 33], we focus on the three-receiver system of Figure 8. The results given simplify easily to the two-receiver case and generalize to more receivers. Note that, since we consider discrete channels, the degraded broadcast channel converses of [47] or of [48], which allow no common information or partial common information, are applicable. In the given broadcast system source coding model, samples of source vector $(U_1, U_2, U_3, U_{12}, U_{23}, U_{13}, U_{123})$ are drawn i.i.d from some distribution $p(u_1, u_2, u_3, u_{12}, u_{23}, u_{13}, u_{123})$. The source description contains components of rates $R_1$, $R_2$, $R_3$, $R_{12}$, $R_{23}$, $R_{13}$, and $R_{123}$. Decoder 1 receives the rate $R_1$, $R_{12}$, $R_{13}$, and $R_{123}$ descriptions and uses them to decode $(U_1, U_{12}, U_{13}, U_{123})$. Decoder 2 receives the rate $R_2$, $R_{12}$, $R_{23}$, and $R_{123}$ descriptions and uses them to decode $(U_2, U_{12}, U_{23}, U_{123})$. Decoder 3 receives the rate $R_3$, $R_{13}$, $R_{23}$, and $R_{123}$ descriptions and uses them to decode $(U_3, U_{13}, U_{23}, U_{123})$. While several receivers decode the common information, each has a different subset of the descriptions with which to decode.

The following theorem proves the optimality of linear broadcast system source codes.

**Theorem 18** *Consider samples of source vector $(U_1, U_2, U_3, U_{12}, U_{23}, U_{13}, U_{123})$ drawn i.i.d according to distribution $p(u_1, u_2, u_3, u_{12}, u_{23}, u_{13}, u_{123})$ on $(\mathbb{F}_2)^7$ and linear broadcast system source codes of rate-$(R_1, R_2, R_3, R_{12}, R_{23}, R_{13}, R_{123})$ and codeword length $n$. For any $s \subseteq \{1, 2, 3, 12, 23, 13, 123\}$, let $u_s = (u_a)_{a \in s}$, and let $(nR)_s = \sum_{a \in s} \lceil nR_a \rceil$. Then for any rates*

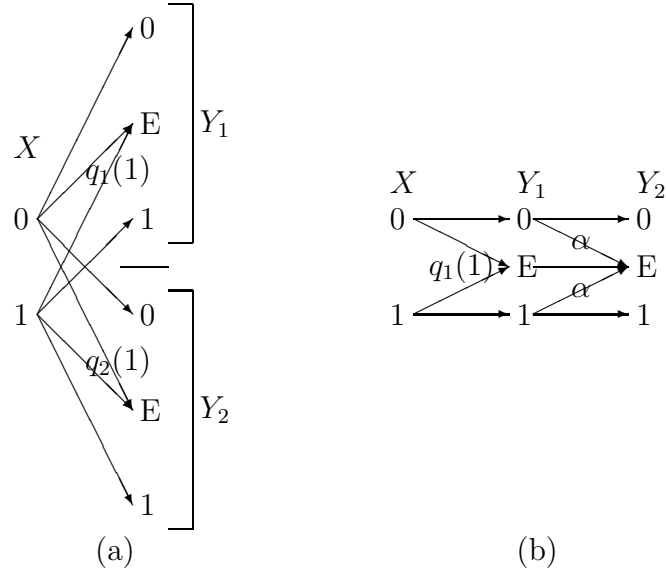

Figure 9: (a) The erasure broadcast channel and (b) a physically degraded channel with the same capacity ($\alpha = (q_2(1) - q_1(1))/(1 - q_1(1))$ and all erasures propagate as erasures).

*satisfying*

$$(nR)_s \geq H(U_s|U_{S_1-s}) \quad \forall \quad s \subseteq S_1 = \{1, 12, 13, 123\}, s \neq \phi$$
$$(nR)_s \geq H(U_s|U_{S_2-s}) \quad \forall \quad s \subseteq S_2 = \{2, 12, 23, 123\}, s \neq \phi$$
$$(nR)_s \geq H(U_s|U_{S_3-s}) \quad \forall \quad s \subseteq S_3 = \{3, 13, 23, 123\}, s \neq \phi$$

*the error probability averaged over the ensemble of linear broadcast system source codes tends to 0 as codeword length tends to $\infty$.*

*Proof:* See Appendix 5. □

## 5.2 Linear Channel Codes for the Erasure Broadcast Channel

We next consider the erasure broadcast channel models shown in Figure 9 (a) and (b). A single channel input is sent to receivers 1 and 2. In the first model, the output at receiver 1 is an erasure with probability $q_1(1)$ and the transmitted value with probability $q_1(0)$; likewise, the output at receiver 2 is an erasure with probability $q_2(1)$ and is otherwise received correctly. Without loss of generality, assume that $q_1(1) \leq q_2(1)$. In this model, erasures are assumed to be independent events. In the model of Figure 9(b), the erasure probabilities for the two receivers are the same, but the erasures are dependent random variables, with all erasures at the first receiver propagating to the second receiver. By [1, Theorem 14.6.1], the capacity of the broadcast channel depends only on the conditional marginal distributions $p(y_1|x)$ and $p(y_2|x)$, thus the capacity of the two channels shown and all channels with the same $p(y_1|x)$ and $p(y_2|x)$ (regardless of the statistical dependencies between erasure events $Z_1$ and $Z_2$) are identical.[5] Note that the elegant and simple converse for degraded BSC broadcast channels of [46], which relies on properties of binary sequences, might be readily extended to our model, albeit without the generality of [47, 48].

---

[5]All channel models considered here assume $Z_1$ and $Z_2$ are independent of the channel input.

### 5.2.1 Capacity of Erasure Broadcast Channel

Lemma 5 proves time-sharing to be optimal for broadcast coding over the given family of channels. The result of Theorem 19, proving the rates achievable by linear broadcast channel codes on the erasure broadcast channel is then immediate by the previous linearity of time-sharing argument. The given bound is optimal for the case of no common information. No converse exists for the case of common information, but the given linear coding achievability results agree with the best known achievability results on the binary erasure channel.

**Lemma 5** *Consider a binary erasure channel with output alphabets $\{0, 1, E\}$ at each of two receivers. The erasure sequences $Z_{1,1}, Z_{1,2}, \ldots$ and $Z_{2,1}, Z_{2,2}, \ldots$ are drawn i.i.d according to distributions $q_1(z_1)$ and $q_2(z_2)$, respectively, where $Z_{i,j} = 1$ denotes an erasure event at receiver $i$ at time $j$. The joint distribution $q(z_1, z_2)$ may be any distribution with the given marginals, but the channel noise is independent of the channel input by assumption. The capacity region for sending independent information to the two receivers is described by*

$$\frac{R_1}{1 - q_1(1)} + \frac{R_2}{1 - q_2(1)} \leq 1.$$

*For any achievable independent information rate pair $(R_1, R_2)$, the rate triple $(R'_1, R'_2, R'_{12}) = (R_1, R_2 - R_0, R_0)$ with common information rate $R'_{12}$ and independent information rates $R'_1$ and $R'_2$ is also achievable for any $R_0 < R_2$.*

*Proof:* See Appendix 5. □

The following theorem shows that random linear channel codes are optimal for the erasure broadcast channel.

**Theorem 19** *Consider an erasure channel with input alphabet $\mathbb{F}_2$ and output alphabets $\{0, 1, E\}$ at each of two receivers. The erasure sequences $Z_{1,1}, Z_{1,2}, \ldots$ and $Z_{2,1}, Z_{2,2}, \ldots$ are drawn i.i.d according to distributions $q_1(z_1)$ and $q_2(z_2)$, respectively, where $Z_{i,j} = 1$ denotes an erasure event at receiver $i$ at time $j$. The joint distribution $q(z_1, z_2)$ may be any distribution with the given marginals, but the channel noise is independent of the channel input by assumption. Let $\{B_n\}_{n=1}^{\infty}$ describe a sequence of channel codes. Each $B_n$ is an $n \times (\lfloor nR_1 \rfloor + \lfloor nR_2 \rfloor)$ matrix with elements chosen i.i.d Bernoulli(1/2). If $R_1/(1 - q_1(1)) + R_2/(1 - q_2(1)) < 1$, then the expected error probability $E[P_e(B_n)] \to 0$ as $n \to \infty$.*

## 5.3 Linear Joint Source-Channel Coding for the Erasure Broadcast Networks

We have seen in the previous subsections that linear source and channel codes are optimal for the erasure broadcast channel. Moreover, Lemma 5 shows that time-sharing is optimal for broadcast coding which establishes that source-channel separation holds for this class of channels. Therefore, the linear source and channel codes can be combined to yield optimal linear joint source-channel codes. This shows the optimality of linear joint source-channel codes for the erasure broadcast channel.

## 5.4 Additive Noise Broadcast Networks

To date, there exist no results to prove the optimality of linear broadcast codes for the additive noise broadcast channel model. In this case, time-sharing is not the optimal solution [1], and direct application of the techniques used in this paper fail to achieve the optimal performance. The stumbling block is that we cannot apply the construction used to build channel input $X$ from the auxiliary random variable $W$ to be decoded by the second receiver. (See the proof of Lemma 5.) In particular, we cannot achieve the appropriate (non-uniform) cross-over probability from the auxiliary random variable to $X$ using an additive signal created by a linear code. In this case, as in Theorem 19, the time-sharing solution is achievable with linear coding. While the time-sharing solution gives a bound on the performance achievable by linear coding, the time-sharing solution is sub-optimal for this problem. Linear coding performance beyond the time-sharing bound may or may not be possible. A possible strategy for trying to move linear codes beyond the time-sharing bound is described in Appendix 5.

# 6 End-to-End Coding and Conclusions

## 6.1 End-to-End Coding

The preceding sections treat the topics of source and channel coding using the tools of linear network coding, bringing previously disparate areas into a common framework. We end by demonstrating that this unification is not only useful in its combination of tasks once treated entirely separately but is in fact crucial to achieving optimal, reliable communication.

Traditional routing techniques rely entirely on repeat and forward strategies for getting a source from its point of origin to its desired destination. The network coding literature demonstrates the failure of that approach in achieving the optimal performance for some simple multi-cast examples [34]. We next demonstrate the failure of the network coding model.

The common network coding model assumes that all sources are independent and all links are noiseless. Implicit in the given model is the assumption that source and channel coding are performed separately from network coding at the edges of the network, so that the internal nodes need only pass along the information to the appropriate receivers. We next demonstrate that source-network separation and channel-network separation both fail. That is, there exist networks for which network coding and source coding must be performed jointly in order to achieve the optimal performance. Likewise, there exist networks for which network coding and channel coding must be performed jointly in order to achieve the optimal performance. We use a sequence of simple examples to prove these results.

**Example 1:** The network of Figure 10 comprises two transmitters and three receivers. Receiver node 1 wishes to receive $U_1$, receiver node 2 wishes to receive $U_2$, and receiver node

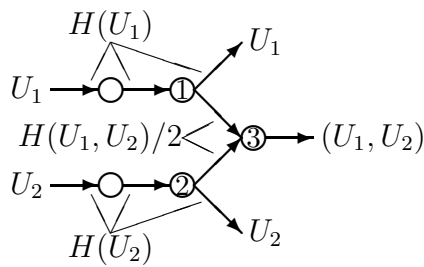

Figure 10: A network for which separation of source and network coding fails.

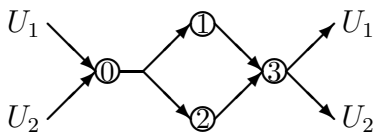

Figure 11: A network for which separation of channel and network coding fails.

3 wishes to receive both $U_1$ and $U_2$. Sources $(U_1, U_2)$ are dependent random variables, with $H(U_1, U_2) < H(U_1) + H(U_2)$. All network links are lossless, and the capacities are noted in the figure. Achieving reliable communication in this example requires the descriptions received by nodes 1 and 2 to be dependent random variables and requires sources $U_1$ and $U_2$ to be re-compressed at nodes 1 and 2, respectively. Thus separation of source coding and network coding fails. □

**Example 2:** Consider the network shown in Figure 11. The channel between node 0 and nodes 1 and 2 is a broadcast erasure channel with independent erasures of probabilities $q_1(1) = q_2(1) = q$. The network between nodes 1 and 2 and node 3 is a multiple access channel without interference. The network coding approach requires labeling each link with its corresponding link capacity. If $R_1$ and $R_2$ are the capacities of the edges to receivers 1 and 2, then $R_1 + R_2$ must be less than $1 - q$ by Theorem 19. The links from node 1 to node 3 and from node 2 to node 3 are both lossless, with capacity 1 bit per channel use. Optimal network coding on the given channel gives a maximal rate of $1 - q$ from the encoder to the decoder. We contrast with the above separated channel and network coding approach an end-to-end coding strategy. In this case, we do not force zero error probability between node 0 and nodes 1 and 2 but instead simply forward the information received by those nodes to the decoder. The capacity of the resulting code is $1 - q^2$ since receiver 3 suffers an erasure only if both node 1 and node 2 receive erasures. □

In addition to illustrating the failure of separate channel and network coding schemes, Example 2 serves as a reminder that general network capacities cannot be proven by breaking the network into canonical elements and solving them independently. Sadly, the strategy given for that example is not always optimal. In particular, the strategy discussed in Example 2 demonstrates that failure to decode at intermediate nodes of the network can yield performance superior to that achieved by decoding at intermediate nodes. Example 3 gives an example that teaches the opposite lesson.

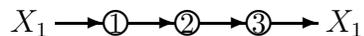

Figure 12: A network for which separation of source and network coding fails. The links between nodes 1 and 2 and nodes 2 and 3 are independent erasure channels with probabilities of erasure $q_1(1)$ and $q_2(1)$, respectively.

**Example 3:** Consider the channel of Figure 12. The links between nodes 1 and 2 and nodes 2 and 3 are independent erasure channels with probabilities of erasure $q_1(1)$ and $q_2(1)$, respectively. If we do not decode at the intermediate node, then the maximal achievable rate from node 1 to node 3 is $(1 - q_1(1))(1 - q_2(1))$. Decoding at node 2 yields maximal achievable rate $\min\{1 - q_1(1), 1 - q_2(1)\} \geq (1 - q_1(1))(1 - q_2(1))$.

The failure of separation in Examples 1 and 2 and the contrasting lessons regarding decoding at intermediate nodes demonstrated by Examples 2 and 3 make the case for the need for end-to-end coding in network environments. The success of the linear coding technique in network coding, source coding, and channel coding suggests that a unified approach that obviates the need for separate routing, compression, and error correction codes may be within reach. In contrast, the failure of separation across canonical network systems seems to present a far greater challenge to optimal code design in networks.

## 6.2 Conclusions

In this paper, we consider networks operating over a common finite field. We show that linear codes are optimal for the point-to-point, multiple access and erasure broadcast networks. We prove that for multiple access networks, source-channel separation holds as long as noise is independent of inputs. We show that separation may fail for binary multiple access networks with input-dependent additive noise. We present an optimal systematic multiple access channel code construction and also provide coding techniques when transmitters are bursty. We show with examples that design for individual network modules may yield poor results when such modules are concatenated, establishing the necessity for end-to-end coding. Thus, we show that it is the lack of decomposability into canonical network modules that is a much greater challenge than the lack of separation between source and channel coding.

# Appendix 1

**Proof of Lemma 3:**
The proof of this lemma uses the analogue of the Darmois-Skitovich theorem for discrete periodic Abelian groups by Fel'dman [45]. Let us proceed by contradiction. Suppose that the $j$th column of $a$ has non-zero elements in positions $i$ and $\hat{i}$ ($\hat{i} \neq i$). Then $V_{\hat{i}}$ and $V_i$ both experience a non-zero contribution from $U_j$. In this case, the independence of $V_{\hat{i}}$ and $V_i$ requires that $p_j$ be a uniform probability mass function, which gives a contradiction. □

**Proof of Theorem 2:**
To accomplish linear channel coding for the erasure channel, we use an $n \times \lfloor nR \rfloor$ linear generator matrix $B_n$ and a conceptually simple non-linear decoder. The linear channel encoder is defined by
$$\gamma(v^{\lfloor nR \rfloor}) = B_n \mathbf{v}.$$

Let $X^n$ denote the channel input and $Y^n$ denote the corrupted channel output. For any $y^n = \mathbf{y}^t \in \{0, 1, \mathsf{E}\}^n$ define the decoder as

$$\delta_n(y^n) = \begin{cases} v^n & \text{if } (B_n \mathbf{v})_i = y_i \text{ for all } i \text{ s.t. } y_i \in \mathbb{F}_2 \\ & \text{and } \nexists \hat{\mathbf{v}} \neq \mathbf{v} \text{ s.t. } (B_n \hat{\mathbf{v}})_i = y_i \text{ for all } i \text{ s.t. } y_i \in \mathbb{F}_2 \\ \hat{V}^{\lfloor nR \rfloor} & \text{otherwise,} \end{cases}$$

where for any $\mathbf{v} \in \mathbb{F}_2^{\lfloor nR \rfloor}$, $(B_n \mathbf{v})_i$ is the $i$th component of the vector $B_n \mathbf{v}$. Again, decoding to $\hat{V}^{\lfloor nR \rfloor}$ denotes a random decoder output.

For the erasure channel, we can immediately decode $Z^n$ from the received string $Y^n$. For any $z^n \in \mathbb{F}_2^n$, define $\mathcal{E}(z^n) = \{\mathbf{e} \in \mathbb{F}_2^n : e_i = z_i \ \forall i \text{ s.t. } z_i = 0\}$. A decoding error occurs if there exists a $\hat{\mathbf{v}} \neq \mathbf{V}$ for which $B_n \mathbf{V} - B_n \hat{\mathbf{v}} = B_n(\mathbf{V} - \hat{\mathbf{v}}) \in \mathcal{E}(Z^n)$, since any such $\hat{\mathbf{v}}$ would be mapped to the same channel output by $Z^n$. For any $z^n$ with $\sum_{i=1}^n z_i = k$, $|\mathcal{E}(z^n)| = 2^k$. Using the definition of the typical set, $z^n \in A_\epsilon^{(n)}$ implies that $\sum_{i=1}^n z_i \leq n(q(1) + \epsilon')$, where $\epsilon' = \epsilon/\log(q(1)/q(0))$. Thus for any fixed $z^n \in A_\epsilon^{(n)}$ and $\mathbf{w}^t \in \mathbb{F}_2^{\lfloor nR \rfloor}$, $\Pr(B_n \mathbf{w} \in \mathcal{E}(z^n)) \leq 2^{-n} 2^{n(q(1)+\epsilon')}$ (since $B_n \mathbf{w}$ is uniformly distributed by the argument in the proof of Theorem 1), giving

$$\begin{aligned} & E[P_e^{(n)}(B_n)] \\ &= E[\Pr(\text{Error} \wedge Z^n \notin A_\epsilon^{(n)}(q))] + E[\Pr(\text{Error} \wedge Z^n \in A_\epsilon^{(n)}(q))] \\ &\leq \epsilon_n + \sum_{v^{\lfloor nR \rfloor}, \hat{v}^{\lfloor nR \rfloor} \in \mathbb{F}_2^{\lfloor nR \rfloor}} \sum_{z^n \in A_\epsilon^{(n)}(q)} p(v^{\lfloor nR \rfloor}) q(z^n) \mathbf{1}(\hat{\mathbf{v}} \neq \mathbf{v}) \Pr(B_n(\mathbf{v} - \hat{\mathbf{v}}) \in \mathcal{E}(z^n)) \\ &\leq \epsilon_n + \sum_{v^{\lfloor nR \rfloor} \in \mathbb{F}_2^{\lfloor nR \rfloor}} \sum_{z^n \in A_\epsilon^{(n)}(q)} p(v^{\lfloor nR \rfloor}) q(z^n) 2^{\lfloor nR \rfloor} 2^{-n} 2^{n(q(1)+\epsilon')} \\ &\leq \epsilon_n + 2^{-n(1-q(1)-\epsilon') + \lfloor nR \rfloor} \end{aligned}$$

for some $\epsilon_n \to 0$. Here $A_\epsilon^{(n)}(p)$ is the typical set for the source distribution and $A_\epsilon^{(n)}(q)$ is the typical set for the noise. The expected error probability decays to zero as $n$ grows without bound provided that $R < 1 - q(1) - \epsilon'$. □

**Proof of Theorem 4:**
The joint source-channel code's encoder is defined by
$$\zeta(u^n) = C_n \mathbf{u}.$$

Denote the random channel input and output by $X^n$ and $Y^n$, respectively. For any $y^n = \mathbf{y}^t \in \{0, 1, \mathsf{E}\}^n$ the decoder is defined by

$$\eta_n(y^n) = \begin{cases} u^n & \text{if } (C_n \mathbf{u})_i = y_i \text{ for all } i \text{ s.t. } y_i \in \mathbb{F}_2 \\ & \text{and } \nexists \hat{\mathbf{u}} \neq \mathbf{u} \text{ s.t. } (C_n \hat{\mathbf{u}})_i = y_i \text{ for all } i \text{ s.t. } y_i \in \mathbb{F}_2 \\ \hat{U}^n & \text{otherwise.} \end{cases}$$

Here, $(C_n\mathbf{u})_i$ denotes the $i$th component of vector $C_n\mathbf{u}$. The error probability for code $C_n$ is

$$P_e(C_n) = \Pr(\eta_n(\zeta_n(Y^n)) \neq U^n),$$

where $U^n$ and $Y^n$ are the random source vector and channel output, respectively. Theorem 4 demonstrates that the expected error probability for a randomly chosen linear code $C_n$ decays to zero as $n$ grows without bound.

Again, we can immediately decode $Z^n$ from the received string $Y^n$, and a decoding error occurs if there exists a $\hat{\mathbf{u}} \neq \mathbf{U}$ for which $C_n(\mathbf{U} - \hat{\mathbf{u}}) \in \mathcal{E}(Z^n)$. Thus

$$\begin{aligned}
E[P_e^{(n)}(C_n)] &= E[\Pr\left(\text{Error} \wedge \left(U^n \notin A_\epsilon^{(n)}(p) \vee Z^n \notin A_\epsilon^{(n)}(q)\right)\right)] \\
&\quad + E[\Pr\left(\text{Error} \wedge U^n \in A_\epsilon^{(n)}(p) \wedge Z^n \in A_\epsilon^{(n)}(q)\right)] \\
&\leq 2\epsilon_n + \sum_{u^n, \hat{u}^n \in A_\epsilon^{(n)}(p)} \sum_{z^n \in A_\epsilon^{(n)}(q)} p(u^n) q(z^n) \mathbf{1}(\hat{\mathbf{u}} \neq \mathbf{u}) \Pr(C_n(\mathbf{u} - \hat{\mathbf{u}}) \in \mathcal{E}(z^n)) \\
&\leq 2\epsilon_n + \sum_{u^n \in A_\epsilon^{(n)}(p)} \sum_{z^n \in A_\epsilon^{(n)}(q)} p(u^n) q(z^n) 2^{n(H(U)+\epsilon)} 2^{-n} 2^{n(q(1)+\epsilon')} \\
&\leq 2\epsilon_n + 2^{-n(1-q(1)-\epsilon'-H(U)-\epsilon)}
\end{aligned}$$

for some $\epsilon_n \to 0$. Thus the expected error probability decays to zero as $n$ grows without bound provided that $H(U) < 1 - q(1) - \epsilon - \epsilon'$. $\square$

**Proof of Theorem 5:**

We now consider a linear joint source-channel code for the binary additive noise channel. Again, given $n \times n$ matrix $C_n$, we define the encoder as

$$\zeta(u^n) = C_n\mathbf{u}.$$

Given random channel input $C_n\mathbf{U}$, the channel output is

$$\mathbf{Y} = C_n\mathbf{U} + \mathbf{Z}.$$

The decoder is

$$\eta_n(y^n) = \begin{cases} u^n & \text{if } u^n \in A_\epsilon^{(n)}(p) \text{ and } \exists z^n \in A_\epsilon^{(n)}(q) \text{ s.t. } C_n\mathbf{u} + \mathbf{z} = \mathbf{y} \\ & \text{and } \nexists (\hat{\mathbf{u}}^n, \hat{\mathbf{z}}^n) \in (A_\epsilon^{(n)}(p) \cap \{\mathbf{u}\}^c) \times A_\epsilon^{(n)}(q) \text{ s.t. } C_n\hat{\mathbf{u}} + \hat{\mathbf{z}} = \mathbf{y} \\ \hat{U}^n & \text{otherwise.} \end{cases}$$

The error probability for code $C_n$ is

$$P_e(C_n) = \Pr(\eta_n(\zeta_n(U^n) + Z^n) \neq U^n).$$

An error occurs if two source sequences are mapped to the same channel input vector or if there exist distinct noise vectors that map distinct channel input vectors to the same channel output. In the first case, $C_n\mathbf{U} = C_n\hat{\mathbf{u}}$ for some $\hat{\mathbf{u}} \neq \mathbf{U}$, and in the second case, $C_n\mathbf{U} + \mathbf{Z} = C_n\hat{\mathbf{u}} + \hat{\mathbf{z}}$ for some $\hat{\mathbf{u}} \neq \mathbf{U}$ and $\hat{\mathbf{z}} \neq \mathbf{Z}$. Restricting our attention to typical source and noise vectors, an error occurs if there exists a $\hat{\mathbf{u}} \in A_\epsilon^{(n)}(p)$ such that $\hat{\mathbf{u}} \neq \mathbf{U}$ and $C_n(\hat{\mathbf{u}} - \mathbf{U}) \in \{\mathbf{0}\} \cup \{\hat{\mathbf{z}} - \mathbf{Z} : \hat{\mathbf{z}} \in A_\epsilon^{(n)}(q)\}$. For any fixed $\mathbf{u} - \hat{\mathbf{u}} \neq \mathbf{0}$ and randomly chosen $C_n$, the coefficients of vector $C_n(\mathbf{u} - \hat{\mathbf{u}})$ are sums of fixed numbers of i.i.d Bernoulli(1/2) values. Thus $\Pr(C_n(\hat{\mathbf{u}} - \mathbf{u}) = \mathbf{w}) = 2^{-n}$ for all $\mathbf{w} \in \mathbb{F}_2^n$, and

$$
\begin{aligned}
&E[P_e^{(n)}(C_n)] \\
&= E[\Pr\left(\text{Error} \wedge \left(U^n \notin A_\epsilon^{(n)}(p) \vee Z^n \notin A_\epsilon^{(n)}(q)\right)\right)] \\
&\quad + E[\Pr\left(\text{Error} \wedge U^n \in A_\epsilon^{(n)}(p) \wedge Z^n \in A_\epsilon^{(n)}(q)\right)] \\
&\leq 2\epsilon_n + \sum_{(u^n,z^n),(\hat{u}^n,\hat{z}^n) \in A_\epsilon^{(n)}(p) \times A_\epsilon^{(n)}(q)} p(u^n)q(z^n)\mathbf{1}(\hat{\mathbf{u}} \neq \mathbf{u})\Pr\left(C_n(\mathbf{u} - \hat{\mathbf{u}}) = \hat{\mathbf{z}} - \mathbf{z}\right) \\
&\leq 2\epsilon_n + \sum_{(u^n,z^n) \in A_\epsilon^{(n)}(p) \times A_\epsilon^{(n)}(q)} p(u^n)q(z^n) 2^{n(H(U)+\epsilon)} 2^{n(H(q)+\epsilon)} 2^{-n} \\
&\leq 2\epsilon_n + 2^{-n(1-H(q)-H(U)-2\epsilon)}
\end{aligned}
$$

for some $\epsilon_n \to 0$. The error probability goes to zero provided that $H(U) < 1 - H(q) - 2\epsilon$. □

# Appendix 2

**Proof of Lemma 4:**
For the multiple access channel with erasures, the NSMA capacity region, $\mathbb{R}_{\mathsf{NSMA}}^{\mathsf{erasure}}$, is

$$\mathbb{R}_{\mathsf{NSMA}}^{\mathsf{erasure}} = \left\{(R_1, R_2) : R_1 + R_2 < 1 - q(1)\right\}.$$

Similarly, the NSMA capacity region for the additive noise multiple access channel, $\mathbb{R}_{\mathsf{NSMA}}^{\mathsf{add-noise}}$, is

$$\mathbb{R}_{\mathsf{NSMA}}^{\mathsf{add-noise}} = \left\{(R_1, R_2) : R_1 + R_2 < 1 - H(Z)\right\}.$$

Hence, for both channels, the NSMA capacity regions are triangles and time-sharing can achieve any point in the region. $\square$

**Proof of Theorems 7 and 8:**
The following argument demonstrates the construction of linear multiple access channel codes from linear channel codes for single-transmitter, single-receiver networks:
Matrix pair $(B_{n,1}, B_{n,2})$ denotes a linear multiple access channel code with encoders

$$\begin{aligned}
\gamma_1(v_1^{\lfloor nR_1 \rfloor}) &= B_{1,n}\mathbf{v}_1 \\
\gamma_2(v_2^{\lfloor nR_2 \rfloor}) &= B_{2,n}\mathbf{v}_2.
\end{aligned}$$

We build matrices $(B_{n,1}, B_{n,2})$ from the linear code for the corresponding single-transmitter, single-receiver channel. Let $\{B_n\}_{n=1}^{\infty}$ be a sequence of rate-$R$ single-transmitter, single-receiver channel codes for the given channel model, then matrix pair $(B_{n,1}^0, B_{n,2}^0) = (B_n, \mathbf{0}_{nR \times n})$ describes a multiple access channel code that achieves rate pair $(R, 0)$. Similarly, matrix pair $(B_{n,1}^1, B_{n,2}^1) = (\mathbf{0}_{nR \times n}, b_n)$ describes a multiple access channel code achieving rate pair $(0, R)$. The multiple access channel code achieving the $(\lambda, 1 - \lambda)$ time-sharing solution between $(R, 0)$ and $(0, R)$ is a linear code with

$$[B_{1,n}^{\lambda}, B_{2,n}^{\lambda}] = \left(\begin{bmatrix} B_{\lambda n} & \mathbf{0}_{\lambda n \times (1-\lambda)nR} \\ \mathbf{0}_{(1-\lambda)n \times \lambda nR} & \mathbf{0}_{(1-\lambda)n \times (1-\lambda)nR} \end{bmatrix}, \begin{bmatrix} \mathbf{0}_{\lambda n \times \lambda nR} & \mathbf{0}_{\lambda n \times (1-\lambda)nR} \\ \mathbf{0}_{(1-\lambda)n \times \lambda nR} & B_{(1-\lambda)n} \end{bmatrix}\right).$$

We decode the first $\lambda n$ channel outputs with the decoder for $B_{\lambda n}$ and the remaining outputs with the decoder for $\beta_{(1-\lambda)n}$. The resulting codes lead immediately to Theorems 7 and 8. While the proofs of Theorems 7 and 8 take slightly different approaches, this difference is not essential. The proof methodology from Theorem 7, which uses direct typical set decoding rather than building a parity-check matrix, can be adapted to the additive noise multiple access channel. $\square$

**Proof of Theorem 9:**
Let us first consider the binary multiple access channel shown in Figure 5(a) with erasures that are independent of the inputs. As seen in Lemma 4, the NSMA capacity is

$$\mathbb{R}_{\mathsf{NSMA}}^{\mathsf{erasure}} = \left\{(R_1, R_2) : R_1 + R_2 < 1 - q(1)\right\}. \tag{18}$$

The three mutual information terms, $I(X_1;Y|X_2)$, $I(X_2;Y|X_1)$ and $I(X_1,X_2;Y)$, are maximized by uniform distribution on $X_1$ and $X_2$. For the same channel, the CMA capacity is

$$\mathbb{R}_{\text{CMA}}^{\text{erasure}} = \left\{ (R_1, R_2) : R_1 + R_2 < 1 - q(1) \right\}, \tag{19}$$

where, the three mutual information terms, $I(X_1;Y|X_2)$, $I(X_2;Y|X_1)$ and $I(X_1,X_2;Y)$, are maximized by making $P(X_1 = i, X_2 = j) = \frac{1}{4}$ for $i, j \in \{0, 1\}$. Combining (18, 19), we obtain

$$\mathbb{R}_{\text{NSMA}}^{\text{erasure}} = \mathbb{R}_{\text{CMA}}^{\text{erasure}}.$$

Hence, by Lemma 1, separation holds. We have thus proved the theorem for the binary multiple access channel with erasures that are independent of the inputs.

Let us now consider the binary multiple access channel shown in Figure 5(b) with noise being independent of the inputs. As seen in Lemma 4, the NSMA capacity is

$$\mathbb{R}_{\text{NSMA}}^{\text{add-noise}} = \left\{ (R_1, R_2) : R_1 + R_2 < 1 - H(Z) \right\}. \tag{20}$$

The three mutual information terms, $I(X_1;Y|X_2)$, $I(X_2;Y|X_1)$ and $I(X_1,X_2;Y)$, are maximized by uniform distribution on $X_1$ and $X_2$. For the same channel, the CMA capacity is

$$\mathbb{R}_{\text{CMA}}^{\text{add-noise}} = \left\{ (R_1, R_2) : R_1 + R_2 < 1 - H(Z) \right\}, \tag{21}$$

where, the three mutual information terms, $I(X_1;Y|X_2)$, $I(X_2;Y|X_1)$ and $I(X_1,X_2;Y)$, are maximized by making $P(X_1 = i, X_2 = j) = \frac{1}{4}$ for $i, j \in \{0, 1\}$. Combining (20, 21), we obtain

$$\mathbb{R}_{\text{NSMA}}^{\text{add-noise}} = \mathbb{R}_{\text{CMA}}^{\text{add-noise}}.$$

Hence, by Lemma 1, separation holds. We have thus proved the theorem for the binary multiple access channel with additive noise that is independent of the inputs. □

**Proof of Theorem 10:**
Theorem 6 specifies the Slepian-Wolf region for the given source as $R_1 > H(U_1|U_2)$, $R_2 > H(U_2|U_1)$, and $R_1 + R_2 > H(U_1, U_2)$. We see from Theorem 9 that source-channel separation holds and the CMA capacity region is the same as the NSMA capacity region. Since (from Lemma 4) the NSMA capacity region for the given channel is $R_1 + R_2 < 1 - q(1)$, the theorem follows. □

**Proof of Theorem 11:**
The proof follows in the same manner as the proof for the multiple access channel with erasures. Here, by Theorem 6, the Slepian-Wolf region for the given source is $R_1 > H(U_1|U_2)$, $R_2 > H(U_2|U_1)$, and $R_1 + R_2 > H(U_1, U_2)$. We see from Theorem 9 that source-channel separation holds and the CMA capacity region is the same as the NSMA capacity region. Since (from Lemma 4) the capacity region for the given channel is $R_1 + R_2 < 1 - H(Z)$, the theorem follows. $\square$

**Proof of Theorem 12:**
Again, we begin by noting the erasure positions in $Y^n$ and using them to reconstruct $Z^n$. A decoding error occurs if there exists a $\hat{\mathbf{u}}_1 \neq \mathbf{U}_1$ for which $C_{1,n}(\mathbf{U}_1 - \hat{\mathbf{u}}_1) \in \mathcal{E}(Z^n)$, a $\hat{\mathbf{u}}_2 \neq \mathbf{U}_2$ for which $C_{2,n}(\mathbf{U}_2 - \hat{\mathbf{u}}_2) \in \mathcal{E}(Z^n)$, or a $\hat{\mathbf{u}}_1 \neq \mathbf{U}_1$ and $\hat{\mathbf{u}}_2 \neq \mathbf{U}_2$ for which $C_{1,n}(\mathbf{U}_1 - \hat{\mathbf{u}}_1) + C_{2,n}(\mathbf{U}_2 - \hat{\mathbf{u}}_2) \in \mathcal{E}(Z^n)$. Thus

$$
\begin{aligned}
&E[P_e^{(n)}(C_{1,n}, C_{2,n})] \\
&= E[\Pr\left(\text{Error} \wedge \left((U_1^n, U_2^n) \notin A_\epsilon^{(n)}(p) \vee Z^n \notin A_\epsilon^{(n)}(q)\right)\right)] \\
&\quad + E[\Pr\left(\text{Error} \wedge (U_1^n, U_2^n) \in A_\epsilon^{(n)}(p) \wedge Z^n \in A_\epsilon^{(n)}(q)\right)] \\
&\leq 2\epsilon_n + \sum_{(u_1^n, u_2^n) \in A_\epsilon^{(n)}(p)} \sum_{z^n \in A_\epsilon^{(n)}(q)} p(u_1^n, u_2^n) q(z^n) \\
&\quad \cdot \left[ \sum_{\hat{u}_1^n \neq u_1^n : (\hat{u}_1^n, u_2^n) \in A_\epsilon^{(n)}(p)} \Pr(C_{1,n}(\mathbf{u}_1 - \hat{\mathbf{u}}_1) \in \mathcal{E}(z^n)) \right.\\
&\quad + \sum_{\hat{u}_2^n \neq u_2^n : (u_1^n, \hat{u}_2^n) \in A_\epsilon^{(n)}(p)} \Pr(C_{2,n}(\mathbf{u}_2 - \hat{\mathbf{u}}_2) \in \mathcal{E}(z^n)) \\
&\quad \left. + \sum_{\hat{u}_1^n \neq u_1^n, \hat{u}_2^n \neq u_2^n : (\hat{u}_1^n, \hat{u}_2^n) \in A_\epsilon^{(n)}(p)} \Pr(C_{1,n}(\mathbf{u}_1 - \hat{\mathbf{u}}_1) + C_{2,n}(\mathbf{u}_2 - \hat{\mathbf{u}}_2) \in \mathcal{E}(z^n)) \right] \\
&\leq 2\epsilon_n + \sum_{(u_1^n, u_2^n) \in A_\epsilon^{(n)}(p)} \sum_{z^n \in A_\epsilon^{(n)}(q)} p(u_1^n, u_2^n) q(z^n) \\
&\quad \cdot \left[ 2^{n(H(U_1|U_2)+\epsilon)} 2^{-n} 2^{n(q(1)+\epsilon')} + 2^{n(H(U_2|U_1)+\epsilon)} 2^{-n} 2^{n(q(1)+\epsilon')} + 2^{n(H(U_1,U_2)+\epsilon)} 2^{-n} 2^{n(q(1)+\epsilon')} \right] \\
&\leq 2\epsilon_n + 2^{-n(1-q(1)-\epsilon'-H(U_1|U_2)-\epsilon)} + 2^{-n(1-q(1)-\epsilon'-H(U_2|U_1)-\epsilon)} + 2^{-n(1-q(1)-\epsilon'-H(U_1,U_2)-\epsilon)}.
\end{aligned}
$$

for some $\epsilon_n \to 0$. Thus the expected error probability decays to zero as $n$ grows without bound provided that $\max\{H(U_1|U_2), H(U_2|U_1), H(U_1, U_2)\} = H(U_1, U_2) < 1 - q(1) - \epsilon - \epsilon'$. $\square$

**Proof of Theorem 13:**
An error occurs if two values of $u_1^n$ are mapped to the same value of $x_1^n$, two values of $u_2^n$ are mapped to the same value of $x_2^n$, or if there exist distinct noise vectors that map distinct source vectors to the same channel output. In the first case, $C_{1,n}\mathbf{U}_1 = C_{1,n}\hat{\mathbf{u}}_1$ for some $\hat{\mathbf{u}}_1 \neq \mathbf{U}_1$; in the second case, $C_{2,n}\mathbf{U}_2 = C_{2,n}\hat{\mathbf{u}}_2$ for some $\hat{\mathbf{u}}_2 \neq \mathbf{U}_2$; and in the third case, $C_{1,n}\mathbf{U}_1 + \mathbf{Z} = C_{1,n}\hat{\mathbf{u}}_1 + \hat{\mathbf{z}}$ for some $\hat{\mathbf{u}}_1 \neq \mathbf{U}_1$ and $\hat{\mathbf{z}} \neq \mathbf{Z}$, $C_{2,n}\mathbf{U}_2 + \mathbf{Z} = C_{2,n}\hat{\mathbf{u}}_2 + \hat{\mathbf{z}}$ for some $\hat{\mathbf{u}}_2 \neq \mathbf{U}_2$ and $\hat{\mathbf{z}} \neq \mathbf{Z}$, or $C_{1,n}\mathbf{U}_1 + C_{2,n}\mathbf{U}_2 + \mathbf{Z} = C_{1,n}\hat{\mathbf{u}}_1 + C_{2,n}\hat{\mathbf{u}}_2 + \hat{\mathbf{z}}$ for some $\hat{\mathbf{u}}_2 \neq \mathbf{U}_2$, $\hat{\mathbf{u}}_2 \neq \mathbf{U}_2$, and $\hat{\mathbf{z}} \neq \mathbf{Z}$. Thus, setting $\mathcal{F}(z^n) = \{\hat{\mathbf{z}} - \mathbf{z} : \hat{\mathbf{z}} \neq \mathbf{z}, \hat{\mathbf{z}}^t \in A_\epsilon^{(n)}(q)\}$ and restricting our attention to typical error sequences, we sum up the error events as: $C_{1,n}(\mathbf{U}_1 - \hat{\mathbf{u}}_1) \in \{\mathbf{0}\} \cup \mathcal{F}(Z^n)$, $C_{2,n}(\mathbf{U}_2 - \hat{\mathbf{u}}_2) \in \{\mathbf{0}\} \cup \mathcal{F}(Z^n)$, and $C_{1,n}(\mathbf{U}_1 - \hat{\mathbf{u}}_1) + C_{2,n}(\mathbf{U}_2 - \hat{\mathbf{u}}_2) \in \mathcal{F}(Z^n)$. We then bound the expected error probability as

$$\begin{aligned}
&E[P_e^{(n)}(C_{1,n}, C_{2,n})] \\
&= E[\Pr\left(\text{Error} \wedge ((U_1^n, U_2^n) \notin A_\epsilon^{(n)}(p) \vee Z^n \notin A_\epsilon^{(n)}(q))\right)] \\
&\quad + E[\Pr\left(\text{Error} \wedge (U_1^n, U_2^n) \in A_\epsilon^{(n)}(p) \wedge Z^n \in A_\epsilon^{(n)}(q)\right)] \\
&\leq 2\epsilon_n + \sum_{(u_1^n, u_2^n) \in A_\epsilon^{(n)}(p)} \sum_{z^n \in A_\epsilon^{(n)}(q)} p(u_1^n, u_2^n) q(z^n) \\
&\quad \cdot \Bigg[ \sum_{\hat{u}_1^n \neq u_1^n : (\hat{u}_1^n, u_2^n) \in A_\epsilon^{(n)}(p)} \Pr(C_{1,n}(\mathbf{u}_1 - \hat{\mathbf{u}}_1) \in \{\mathbf{0}\} \cup \mathcal{F}(z^n)) \\
&\quad + \sum_{\hat{u}_2^n \neq u_2^n : (u_1^n, \hat{u}_2^n) \in A_\epsilon^{(n)}(p)} \Pr(C_{2,n}(\mathbf{u}_2 - \hat{\mathbf{u}}_2) \in \{\mathbf{0}\} \cup \mathcal{F}(z^n)) \\
&\quad + \sum_{\hat{u}_1^n \neq u_1^n, \hat{u}_2^n \neq u_2^n : (\hat{u}_1^n, \hat{u}_2^n) \in A_\epsilon^{(n)}(p)} \Pr(C_{1,n}(\mathbf{u}_1 - \hat{\mathbf{u}}_1) + C_{2,n}(\mathbf{u}_2 - \hat{\mathbf{u}}_2) \in \mathcal{F}(z^n)) \Bigg] \\
&\leq 2\epsilon_n + \sum_{(u_1^n, u_2^n) \in A_\epsilon^{(n)}(p)} \sum_{z^n \in A_\epsilon^{(n)}(q)} p(u_1^n, u_2^n) q(z^n) \Big[ 2^{n(H(U_1|U_2)+\epsilon)} 2^{-n} 2^{n(H(Z)+\epsilon)} \\
&\quad + 2^{n(H(U_2|U_1)+\epsilon)} 2^{-n} 2^{n(H(Z)+\epsilon)} + 2^{n(H(U_1,U_2)+\epsilon)} 2^{-n} 2^{n(H(Z)+\epsilon)} \Big] \\
&\leq 2\epsilon_n + 2^{-n(1-H(Z)-H(U_1|U_2)-2\epsilon)} + 2^{-n(1-H(Z)-H(U_2|U_1)-2\epsilon)} + 2^{-n(1-H(Z)-H(U_1,U_2)-2\epsilon)}
\end{aligned}$$

for some $\epsilon_n \to 0$. Thus the expected error probability decays to zero as $n$ grows without bound provided that $\max\{H(U_1|U_2), H(U_2|U_1), H(U_1, U_2)\} = H(U_1, U_2) < 1 - H(Z) - 2\epsilon$.
□

## Appendix 3

**Proof of Theorems 14 and 15:**
Let us consider a noisy multiple access channel where two transmitters transmit binary $\{0,1\}$ symbols, $X_1$ and $X_2$, to a single receiver. The received symbol $Y$ is also binary $\{0,1\}$. Binary additive noise $Z$ is allowed to depend on the input symbols being transmitted and has the distribution: $q_{ij} = Pr(Z = 1|X_1 = i, X_2 = j)$ for $i, j \in 0, 1$. Define $P_{ij} = Pr(X_1 = i, X_2 = j)$ for $i, j \in 0, 1$, $p_1 = Pr(X_1 = 0)$ and $p_2 = Pr(X_2 = 0)$. Let us define the function $\mathcal{H}(.)$ as

$$\mathcal{H}(q) = -q \log_2(q) - (1-q) \log_2(1-q) \text{ for } q \geq 0$$

and $\alpha_{00} = 1 - q_{00}$, $\alpha_{01} = q_{01}$, $\alpha_{10} = q_{10}$, $\alpha_{11} = 1 - q_{11}$. Since $q_{ij}$ are probabilities, $\alpha_{ij} \in [0, 1]$ for $i \in \{0, 1\}$. Note that $\alpha_{ij}$ characterizes a particular multiple access channel.

We compute $R_{sum}^{NSMA}(\alpha_{00}, \alpha_{01}, \alpha_{10}, \alpha_{11})$ and $R_{sum}^{CMA}(\alpha_{00}, \alpha_{01}, \alpha_{10}, \alpha_{11})$ as

$$R_{sum}^{CMA}(\alpha_{00}, \alpha_{01}, \alpha_{10}, \alpha_{11}) = \max_{P_{00}, P_{01}, P_{10}, P_{11}} R'_{CMA}(\alpha_{00}, \alpha_{01}, \alpha_{10}, \alpha_{11}, P_{00}, P_{01}, P_{10}, P_{11}), \quad (22)$$

where

$$R'_{CMA}(\alpha_{00}, \alpha_{01}, \alpha_{10}, \alpha_{11}, P_{00}, P_{01}, P_{10}, P_{11}) = \mathcal{H}[P_{00}\alpha_{00} + P_{01}\alpha_{01} + P_{10}\alpha_{10} + P_{11}\alpha_{11}]$$
$$- P_{00}\mathcal{H}(\alpha_{00}) - P_{01}\mathcal{H}(\alpha_{01}) - P_{10}\mathcal{H}(\alpha_{10}) - P_{11}\mathcal{H}(\alpha_{11}), \quad (23)$$

and

$$R^{NSMA}_{sum}(\alpha_{00}, \alpha_{01}, \alpha_{10}, \alpha_{11}) = \max_{p_1, p_2} R'_{NSMA}(\alpha_{00}, \alpha_{01}, \alpha_{10}, \alpha_{11}, p_1, p_2), \quad (24)$$

where

$$R'_{NSMA}(\alpha_{00}, \alpha_{01}, \alpha_{10}, \alpha_{11}, p_1, p_2)$$
$$= \mathcal{H}[p_1 p_2 \alpha_{00} + p_1(1-p_2)\alpha_{01} + p_2(1-p_1)\alpha_{10} + (1-p_1)(1-p_2)\alpha_{11}]$$
$$- p_1 p_2 \mathcal{H}(\alpha_{00}) - p_1(1-p_2)\mathcal{H}(\alpha_{01}) - p_2(1-p_1)\mathcal{H}(\alpha_{10}) - (1-p_1)(1-p_2)\mathcal{H}(\alpha_{11})(25)$$

The difference between the CMA and NSMA multiple access sum capacities, $G(\alpha_{00}, \alpha_{01}, \alpha_{10}, \alpha_{11})$, is

$$G(\alpha_{00}, \alpha_{01}, \alpha_{10}, \alpha_{11}) = R^{CMA}_{sum}(\alpha_{00}, \alpha_{01}, \alpha_{10}, \alpha_{11}) - R^{NSMA}_{sum}(\alpha_{00}, \alpha_{01}, \alpha_{10}, \alpha_{11}).$$

### A.3.1 An example where the CMA and NSMA capacity regions differ

Consider a channel parameterized by $\alpha_{00} = 0$, $\alpha_{01} = 0.5$, $\alpha_{10} = 0.5$ and $\alpha_{11} = 1$. This choice of conditional noise probabilities makes the noise input-dependent. We compute from (22-25):

$$R^{CMA}_{sum}(0, 0.5, 0.5, 1) = 1,$$
$$R^{NSMA}_{sum}(0, 0.5, 0.5, 1) = \frac{1}{2},$$
$$\Rightarrow G(0, 0.5, 0.5, 1) = \frac{1}{2}.$$

Now, we have a channel where the cooperative and separate sum capacities are unequal. This example shows that when noise is allowed to depend on the channel inputs, the NSMA and CMA capacity regions may not be the same. Under this scenario, separation may fail for some source-channel pairs.

### A.3.2 Maximum difference between the CMA and NSMA sum capacities

We now find the maximum difference between the CMA and NSMA sum capacities for binary multiple access channels with additive noise. For this, we need to evaluate

$$\max_{\alpha_{00}, \alpha_{01}, \alpha_{10}, \alpha_{11} \in [0,1]} G(\alpha_{00}, \alpha_{01}, \alpha_{10}, \alpha_{11}).$$

Henceforth, we refer to "sum capacity" as "capacity" for brevity.

**Characteristics of cooperative capacity achieving joint input distribution**
Let us establish the characteristics of the joint input distribution that achieves the cooperative capacity. If $(p_1, p_2)$ achieves the separate capacity and $(P_{00}, P_{01}, P_{10}, P_{11})$ achieves the cooperative capacity of a channel, then the separate and cooperative capacities are the same, $R_{sum}^{NSMA}(\alpha_{00}, \alpha_{01}, \alpha_{10}, \alpha_{11}) = R_{sum}^{CMA}(\alpha_{00}, \alpha_{01}, \alpha_{10}, \alpha_{11})$, for all $\alpha_{00}, \alpha_{01}, \alpha_{10}, \alpha_{11} \in [0, 1]$ if and only if

$$p_1 p_2 = P_{00}, \tag{26}$$
$$p_1(1 - p_2) = P_{01}, \tag{27}$$
$$p_2(1 - p_1) = P_{10}, \tag{28}$$
$$(1 - p_1)(1 - p_2) = P_{11}, \tag{29}$$

by (22-25). For (26-29) to hold, we need

$$P_{11} P_{00} = P_{01} P_{10}. \tag{30}$$

Thus, for any channel, whenever the joint input distribution achieving the CMA capacity obeys (30), the cooperative and separate capacities are the same.

**Maximizing the cooperative mutual information**
We next prove Lemma 7 that identifies the joint input distribution that achieves the CMA capacity for an arbitrary binary multiple access channel. The following definitions and lemma are useful for that proof. Define $\alpha_{\min} = \min\{\alpha_{00}, \alpha_{01}, \alpha_{10}, \alpha_{11}\}$, $\alpha_{\max} = \max\{\alpha_{00}, \alpha_{01}, \alpha_{10}, \alpha_{11}\}$ and $\alpha_1, \alpha_2 \in \{\alpha_{00}, \alpha_{01}, \alpha_{10}, \alpha_{11}\} - \{\alpha_{\min}, \alpha_{\max}\}$. Therefore, $\alpha_{\min}$ and $\alpha_{\max}$ are the smallest and largest $\alpha_{ij}$, respectively, where $i, j \in \{0, 1\}$ and $\alpha_{\min} \leq \alpha_1, \alpha_2 \leq \alpha_{\max}$.

**Lemma 6** *There exists $p \in [0, 1]$, such that*

$$R'_{CMA}(\alpha_{\min}, \alpha_1, \alpha_2, \alpha_{\max}, p, 0, 0, 1 - p) \geq R'_{CMA}(\alpha_{\min}, \alpha_1, \alpha_2, \alpha_{\max}, p'_{\min}, p'_1, p'_2, p'_{\max})$$

*where $(p'_{\min}, p'_1, p'_2, p'_{\max})$ specifies a joint input probability distribution.*

*Proof:* Choose $p$ such that

$$p\alpha_{\min} + (1 - p)\alpha_{\max} = p'_{\min}\alpha_{\min} + p'_1\alpha_1 + p'_2\alpha_2 + p'_{\max}\alpha_{\max}. \tag{31}$$

This is possible, since $p \in [0, 1]$. Now, using (23), the lemma holds if

$$p\mathcal{H}(\alpha_{\min}) + (1 - p)\mathcal{H}(\alpha_{\max}) \leq p'_{min}\mathcal{H}(\alpha_{\min}) + p'_1\mathcal{H}(\alpha_1) + p'_2\mathcal{H}(\alpha_2) + p'_{max}\mathcal{H}(\alpha_{\max}). \tag{32}$$

Solving (31,32),

$$\begin{aligned} 0 &\leq p'_1\left[\mathcal{H}(\alpha_1) - \mathcal{H}(\alpha_{\max}) + \frac{\alpha_1 - \alpha_{\max}}{\alpha_{\max} - \alpha_{\min}}\{\mathcal{H}(\alpha_{\min}) - \mathcal{H}(\alpha_{\max})\}\right] \\ &+ p'_2\left[\mathcal{H}(\alpha_2) - \mathcal{H}(\alpha_{\max}) + \frac{\alpha_2 - \alpha_{\max}}{\alpha_{\max} - \alpha_{\min}}\{\mathcal{H}(\alpha_{\min}) - \mathcal{H}(\alpha_{\max})\}\right], \end{aligned} \tag{33}$$

is a necessary and sufficient condition for the lemma to hold. Since $\mathcal{H}(.)$ is a concave function,

$$0 \leq \mathcal{H}(\alpha_1) - \mathcal{H}(\alpha_{\max}) + \frac{\alpha_1 - \alpha_{\max}}{\alpha_{\max} - \alpha_{\min}} \left[ \mathcal{H}(\alpha_{\min}) - \mathcal{H}(\alpha_{\max}) \right],$$

$$0 \leq \mathcal{H}(\alpha_2) - \mathcal{H}(\alpha_{\max}) + \frac{\alpha_2 - \alpha_{\max}}{\alpha_{\max} - \alpha_{\min}} \left[ \mathcal{H}(\alpha_{\min}) - \mathcal{H}(\alpha_{\max}) \right],$$

since $\alpha_1, \alpha_2 \in [\alpha_{\min}, \alpha_{\max}]$. Moreover, $p_1', p_2' \geq 0$ which implies that (33) holds. The proof is now complete. $\square$

We now prove Lemma 7. Let us define a function $\mathsf{Ind}(.)$ that extracts the indices of its argument. For example $\mathsf{Ind}(\alpha_{ij}) = (i, j)$.

**Lemma 7** $R'_{CMA}(\alpha_{\min}, \alpha_1, \alpha_2, \alpha_{\max}, P_{00}, P_{01}, P_{10}, P_{11})$ *is maximized by the joint probability distribution* $(p_{\min}, 0, 0, 1 - p_{\min})$ *where*

$$p_{\min} = P_{ij}$$

*for the indices* $(i, j)$ *such that* $\alpha_{\min} = \alpha_{ij}$.

*Proof:* For a joint distribution $(p'_{\min}, p_1', p_2', p'_{\max})$, we have, from (22),

$$R^{CMA}_{sum}(\alpha_{\min}, \alpha_1, \alpha_2, \alpha_{\max}) = \max_{p'_{\min}, p_1', p_2', p'_{\max}} R'_{CMA}(\alpha_{\min}, \alpha_1, \alpha_2, \alpha_{\max}, p'_{\min}, p_1', p_2', p'_{\max}),$$

and $\alpha_{\min} \leq \alpha_1, \alpha_2 \leq \alpha_{\max}$. Using Lemma 6, we have

$$R^{CMA}_{sum}(\alpha_{\min}, \alpha_1, \alpha_2, \alpha_{\max}) = \max_{p \in [0,1]} R'_{CMA}(\alpha_{\min}, \alpha_1, \alpha_2, \alpha_{\max}, p, 0, 0, 1 - p). \quad (34)$$

Let (34) be maximized at $p = q^*$. Note that $q^*$ multiplies $\alpha_{\min}$ and $(1 - q^*)$ multiplies $\alpha_{\max}$. We define $p_{\min} = q^*$. Thus, the cooperative mutual information is maximized by the probability distribution $(p_{\min}, 0, 0, 1 - p_{\min})$. The proof is now complete. $\square$

Since $p_{\min}$ multiplies $\alpha_{\min}$ and $1 - p_{\min}$ multiplies $\alpha_{\max}$, the following corollary follows:

**Corollary 2** $R^{CMA}_{sum}(\alpha_{00}, \alpha_{01}, \alpha_{10}, \alpha_{11})$ *depends only on* $\alpha_{\min}$ *and* $\alpha_{\max}$.

For a channel where $\alpha_{ij} = \alpha_{i',j'}$ for $i, j, i', j' \in \{0, 1\}$, there is more than one input probability distribution that achieves the cooperative capacity. Equation (30) must hold for at least one of these distributions for the cooperative and separate capacities to be the same. Hence, while optimizing, when we have more that one choice, we choose $p_{\min}$ such that (30) holds. If (30) does not hold for any of the choices, then the CMA capacity is strictly greater than the NSMA capacity.

**Probability that the cooperative and separate capacities are unequal**
We saw in Corollary 2 that $R^{CMA}_{sum}(\alpha_{00}, \alpha_{01}, \alpha_{10}, \alpha_{11})$ depends only on $\alpha_{\min}$ and $\alpha_{\max}$. This implies that *only* two of the cooperative capacity achieving joint input probabilities, $P_{ij}$, are non-zero. Thus, (30) does not hold when either $P_{00} = P_{11} = 0$ or $P_{01} = P_{10} = 0$. The former condition occurs when $(\alpha_{\min}, \alpha_{\max}) \in \{(\alpha_{01}, \alpha_{10}), (\alpha_{10}, \alpha_{01})\}$ and the latter occurs when $(\alpha_{\min}, \alpha_{\max}) \in \{(\alpha_{00}, \alpha_{11}), (\alpha_{11}, \alpha_{00})\}$. Combining the two, we see that the CMA capacity is strictly larger than the NSMA capacity whenever

$$(\alpha_{\min}, \alpha_{\max}) \in \{(\alpha_{00}, \alpha_{11}), (\alpha_{11}, \alpha_{00}), (\alpha_{01}, \alpha_{10}), (\alpha_{10}, \alpha_{01})\}.$$

Therefore, the separate and cooperative capacities are unequal *if and only if* any one of the following events occur

$$\mathcal{E}_1 : \alpha_{00} < \alpha_{01}, \alpha_{10} < \alpha_{11} \qquad \mathcal{E}_2 : \alpha_{11} < \alpha_{01}, \alpha_{10} < \alpha_{00},$$
$$\mathcal{E}_3 : \alpha_{01} < \alpha_{00}, \alpha_{11} < \alpha_{10} \qquad \mathcal{E}_4 : \alpha_{10} < \alpha_{00}, \alpha_{11} < \alpha_{01}.$$

Since, the four events are disjoint, the probability that capacities are unequal for a channel picked randomly from the ensemble of all such channels, $P_{\text{unequal}}$, is

$$P_{\text{unequal}} = \sum_{i=1}^{4} \Pr(\mathcal{E}_i).$$

If $\alpha_{00}, \alpha_{01}, \alpha_{10}$ and $\alpha_{11}$ are independent and identically distributed in $[0,1]$,

$$\Pr(\mathcal{E}_i) = \frac{1}{12} \text{ for } i \in \{1, 2, 3, 4\}.$$

Thus,

$$P_{\text{unequal}} = \frac{1}{3}.$$

We have thus proved Theorem 15.

**Maximum difference between the cooperative and separate capacities**
We have seen that the cooperative and separate capacities are not the same when

$$(\alpha_{\min}, \alpha_{\max}) \in \{(\alpha_{00}, \alpha_{11}), (\alpha_{11}, \alpha_{00}), (\alpha_{01}, \alpha_{10}), (\alpha_{10}, \alpha_{01})\}.$$

We classify channels for which this happens into two types. Type 1 channels satisfy $(\alpha_{\min}, \alpha_{\max}) \in \{(\alpha_{00}, \alpha_{11}), (\alpha_{11}, \alpha_{00})\}$. Type 2 channels satisfy $(\alpha_{\min}, \alpha_{\max}) \in \{(\alpha_{01}, \alpha_{10}), (\alpha_{10}, \alpha_{01})\}$. Consider a type 2 channel, $C^2$, parameterized by $(\alpha_{00}^{C^2}, \alpha_{01}^{C^2}, \alpha_{10}^{C^2}, \alpha_{11}^{C^2})$. Now, consider another channel, $C^*$, whose parameters are $(\alpha_{00}^{C^*}, \alpha_{01}^{C^*}, \alpha_{10}^{C^*}, \alpha_{11}^{C^*})$ such that

$$\alpha_{00}^{C^*} = \alpha_{01}^{C^2},$$
$$\alpha_{01}^{C^*} = \alpha_{00}^{C^2},$$
$$\alpha_{10}^{C^*} = \alpha_{11}^{C^2},$$
$$\alpha_{11}^{C^*} = \alpha_{10}^{C^2}.$$

Thus, $(\alpha_{\min}^{C^*}, \alpha_{\max}^{C^*}) \in \{(\alpha_{00}^{C^*}, \alpha_{11}^{C^*}), (\alpha_{11}^{C^*}, \alpha_{00}^{C^*})\}$ and this new channel is of type 1. Therefore, we define $C^* \triangleq C^1$. (We use the superscript to designate the channel type.) Note that,

$$(\alpha_{\min}^{C^2}, \alpha_{\max}^{C^2}) = (\alpha_{\min}^{C^1}, \alpha_{\max}^{C^1}).$$

Since the cooperative capacity depends only on the highest and lowest $\alpha_{ij}$,

$$R_{sum}^{CMA}(\alpha_{00}^{C^1}, \alpha_{01}^{C^1}, \alpha_{10}^{C^1}, \alpha_{11}^{C^1}) = R_{sum}^{CMA}(\alpha_{00}^{C^2}, \alpha_{01}^{C^2}, \alpha_{10}^{C^2}, \alpha_{11}^{C^2}). \quad (35)$$

Moreover, if a particular separate capacity is achieved for $C^2$ by a probability distribution $(p_1, p_2)$, the same separate capacity can be achieved for $C^1$ by a probability distribution $(p_1, 1 - p_2)$. Thus,

$$R_{sum}^{NSMA}(\alpha_{00}^{C^1}, \alpha_{01}^{C^1}, \alpha_{10}^{C^1}, \alpha_{11}^{C^1}) = R_{sum}^{NSMA}(\alpha_{00}^{C^2}, \alpha_{01}^{C^2}, \alpha_{10}^{C^2}, \alpha_{11}^{C^2}). \quad (36)$$

From (35,36), we have

$$G(\alpha_{00}^{C^1}, \alpha_{01}^{C^1}, \alpha_{10}^{C^1}, \alpha_{11}^{C^1}) = G(\alpha_{00}^{C^2}, \alpha_{01}^{C^2}, \alpha_{10}^{C^2}, \alpha_{11}^{C^2}).$$

Therefore, for every type 2 channel, there exists a type 1 channel with the same loss in capacity. Hence, while finding the maximum loss, we can confine our analysis to type 1 channels.

Consider two type 1 channels, $C_1^1$ and $C_2^1$, parameterized by $(\alpha_{00}, \alpha_{01}, \alpha_{10}, \alpha_{11})$ and $(\alpha_{00}, 0.5, 0.5, \alpha_{11})$, respectively. As both $C_1^1$ and $C_2^1$ have the same $(\alpha_{\min}, \alpha_{\max})$, we have from Corollary 2,

$$R_{sum}^{CMA}(\alpha_{00}, \alpha_{01}, \alpha_{10}, \alpha_{11}) = R_{sum}^{CMA}(\alpha_{00}, 0.5, 0.5, \alpha_{11}). \quad (37)$$

Since, the noise entropy for $C_2^2$ cannot exceed the noise entropy for $C_2^1$, i.e.,

$$\mathcal{H}_{C_1^1}(Z|X_1 = 0, X_2 = 1) \le \mathcal{H}_{C_2^1}(Z|X_1 = 0, X_2 = 1) = 1,$$
$$\mathcal{H}_{C_1^1}(Z|X_1 = 1, X_2 = 0) \le \mathcal{H}_{C_2^1}(Z|X_1 = 1, X_2 = 0) = 1,$$

we have

$$R_{sum}^{NSMA}(\alpha_{00}, \alpha_{01}, \alpha_{10}, \alpha_{11}) \ge R_{sum}^{NSMA}(\alpha_{00}, 0.5, 0.5, \alpha_{11}). \quad (38)$$

Combining (37) and (38) gives

$$G(\alpha_{00}, \alpha_{01}, \alpha_{10}, \alpha_{11}) \le G(\alpha_{00}, 0.5, 0.5, \alpha_{11}). \quad (39)$$

Thus, in finding the largest loss, it is sufficient to focus on channels parameterized by $(\alpha, 0.5, 0.5, \beta)$ for $\alpha, \beta \in [0, 1]$. We define

$$R_{sum}^{NSMA}(\alpha, \beta) \triangleq R_{sum}^{NSMA}(\alpha, 0.5, 0.5, \beta),$$
$$R_J(\alpha, \beta) \triangleq R_J(\alpha, 0.5, 0.5, \beta),$$
$$G(\alpha, \beta) \triangleq G(\alpha, 0.5, 0.5, \beta).$$

Therefore, the cooperative capacity given by (22,23) becomes

$$R_{sum}^{CMA}(\alpha, \beta) = \max_{P_{00} \in [0,1]} \mathcal{H}[P_{00}\alpha + (1 - P_{00})\beta] - P_{00}\mathcal{H}(\alpha) - (1 - P_{00})\mathcal{H}(\beta). \quad (40)$$

The maximum of (40) occurs at

$$P_{00}^* = \frac{\beta}{\beta - \alpha} - \frac{1}{(\beta - \alpha)(1 + \exp(\phi))},$$

where
$$\phi = \frac{\mathcal{H}(\beta) - \mathcal{H}(\alpha)}{\beta - \alpha}.$$

Substituting, we obtain

$$R_{sum}^{CMA}(\alpha, \beta) \qquad (41)$$
$$= \mathcal{H}\left(\frac{1}{1 + \exp(\phi)}\right) + \phi\beta - \mathcal{H}(\beta) - \frac{\phi}{1 + \exp(\phi)}.$$

The sum rate achievable by separate source-channel coding given by (24,25) is

$$R_{sum}^{NSMA}(\alpha, \beta) =$$
$$\max_{p_1, p_2} \mathcal{H}[p_1 p_2(\alpha + \beta - 1) + (p_1 + p_2)(0.5 - \beta) + \beta]$$
$$- p_1 p_2[\mathcal{H}(\alpha) + \mathcal{H}(\beta) - 2] - (p_1 + p_2)[1 - \mathcal{H}(\beta)] - \mathcal{H}(\beta).$$

Computing $R_{sum}^{NSMA}(\alpha, \beta)$ is difficult in general since the maximizing probability distribution is hard to evaluate. Hence, let us define $R_{NSMA}^{D}(\alpha, \beta)$ as the sum rate due to separate source-channel coding where $(p_1, p_2)$ are chosen to minimize the Euclidian distance between points $(P_{00}^*, 1 - P_{00}^*)$ and $(p_1 p_2, (1 - p_1)(1 - p_2))$. We further impose the constraint that $p_1 = p_2 = p$. By using a probability distribution that minimizes Euclidian distance, we have a lower separate capacity and an upper bound on the difference between the cooperative and separate capacity. Later, we show with an example that the bound is achieved. Thus, minimizing squared Euclidian distance,

$$p^* = \arg\min_p[(P_{00}^* - p^2)^2 + ((1 - P_{00}^*) - (1 - p)^2)^2],$$

and

$$R_{NSMA}^{D}(\alpha, \beta)$$
$$= \mathcal{H}[p^{*2}\alpha + p^*(1 - p^*) + (1 - p^*)^2 \beta] - p^{*2}\mathcal{H}(\alpha) - 2p^*(1 - p^*) - (1 - p^*)^2\mathcal{H}(\beta). \quad (42)$$

Define $G_D(\alpha, \beta)$ as

$$G_D(\alpha, \beta) = R_{sum}^{CMA}(\alpha, \beta) - R_{NSMA}^{D}(\alpha, \beta). \qquad (43)$$

The probability distribution that minimizes Euclidian distance and assumes $p_1 = p_2$ *cannot* yield a mutual information higher than the separate capacity. Therefore,

$$R_{NSMA}^{D}(\alpha, \beta) \le R_{sum}^{NSMA}(\alpha, \beta),$$

and

$$G(\alpha, \beta) \le G_D(\alpha, \beta). \qquad (44)$$

Combining (39,44) we find that for every $\alpha_{00}, \alpha_{01}, \alpha_{10}, \alpha_{11} \in [0, 1]$, there exists $\alpha, \beta \in [0, 1]$ such that

$$G(\alpha_{00}, \alpha_{01}, \alpha_{10}, \alpha_{11}) \le G_D(\alpha, \beta). \qquad (45)$$

**Minimizing Squared Euclidian Distance**

Let

$$d^2 = (P_{00}^* - p^2)^2 + ((1 - P_{00}^*) - (1-p)^2).$$

Setting

$$\frac{\partial d^2}{\partial p} = 0,$$

we obtain

$$2p^3 - 3p^2 + 2p - P_{00}^* = 0.$$

We have three roots for this polynomial. Two are complex and should be omitted. The real root is given by

$$p^* = \frac{1}{2} - \frac{1}{2^{2/3} \sqrt[3]{108 P_{00}^* - 54 + \sqrt{108 + (108 P_{00}^* - 54)^2}}} + \frac{\sqrt[3]{108 P_{00}^* - 54 + \sqrt{108 + (108 P_{00}^* - 54)^2}}}{2^{1/3} 6}. \tag{46}$$

From (41,42,43,46), we have the explicit expression for $G_D(\alpha, \beta)$. We need to find the maximum value of $G_D(\alpha, \beta)$ over the unit square $\alpha, \beta \in [0, 1]$. We have the following lemma:

**Lemma 8**

$$\max_{\alpha, \beta \in [0,1]} G_D(\alpha, \beta) = G_D(0, 1) = G_D(1, 0) = \frac{1}{2}.$$

*Proof:* For $\alpha, \beta \in (0, 1)$, we have

$$\frac{\partial G_D(\alpha, \beta)}{\partial \alpha} \neq 0,$$
$$\frac{\partial G_D(\alpha, \beta)}{\partial \beta} \neq 0.$$

Therefore, no critical points of $G_D(\alpha, \beta)$ lie inside the square region and the maximum values of $G_D(\alpha, \beta)$ must lie on the sides of the square. We obtain the following properties of $G_D(\alpha, \beta)$ over the four sides of the square:

$$\alpha = 0, \beta \in [0,1]: \quad \frac{\partial G_D(0, \beta)}{\partial \beta} > 0$$
$$\beta = 1, \alpha \in [0,1]: \quad \frac{\partial G_D(\alpha, 1)}{\partial \alpha} < 0$$
$$\alpha = 1, \beta \in [0,1]: \quad \frac{\partial G_D(1, \beta)}{\partial \beta} < 0$$
$$\beta = 0, \alpha \in [0,1]: \quad \frac{\partial G_D(\alpha, 0)}{\partial \alpha} > 0$$

Therefore, $G_D(\alpha, \beta)$ takes its *maximum* values at

$$(\alpha = 0, \beta = 1),$$
$$(\alpha = 1, \beta = 0).$$

Using (41,42,43,46), we evaluate $G_D(.,.)$ at these points as

$$G_D(0,1) = G_D(1,0) = \frac{1}{2}.$$

This completes the proof. □

Using Lemma 8 and (45), we have for $\alpha_{00}, \alpha_{01}, \alpha_{10}, \alpha_{11} \in [0,1]$,

$$G(\alpha_{00}, \alpha_{01}, \alpha_{10}, \alpha_{11}) \leq \frac{1}{2}. \tag{47}$$

**Tightness of the bound**

We now show that the bound on $G(\alpha_{00}, \alpha_{01}, \alpha_{10}, \alpha_{11})$ is achieved. For the channel considered in section A.3.1, we have

$$G(0, 0.5, 0.5, 1) = \frac{1}{2}.$$

Now, we have a channel whose CMA capacity is greater than the NSMA capacity by the bound in (47). The bound in (47) is thus achieved.

Therefore, for noisy multiple access binary additive noise channels, the maximum difference between the CMA and NSMA sum capacities is $\frac{1}{2}$ bit per channel use. This completes the proof of Theorem 14. □

# Appendix 4

## A.4.1 Code construction

By definition, $l_a \geq n_a$ and $l_b \geq n_b$. There exist simple codes with $l_a = l_b = n_a + n_b$ for which $v_a = n_a$ and $v_b = n_b$, thus all elements can be recovered after multiple access interference at the receiver. The region $ABCD$ shown in Figure 13 illustrates these constraints. All achievable points outside $ABCD$ have a lower code rate since the number of received bits remains the same while $l_a$ and $l_b$ increase. Hence, we confine our analysis to the region $ABCD$.

In this region $n_a \leq l_a \leq n_a + n_b$ and $n_b \leq l_b \leq n_a + n_b$, which makes $0 \leq m_a \leq n_b$, $0 \leq m_b \leq n_a$, $0 \leq v_a \leq n_a$ and $0 \leq v_b \leq n_b$.

Let $W_a = (TL_a)$ and $W_b = (TL_b)$. We define a *1row* as a row vector having only one non-zero bit. Since $\vec{R}$ contains $v_a$ bits of $\vec{a}$ and $v_b$ bits of $\vec{b}$, $W_a$ should be a $(v_a + v_b) \times n_a$ size matrix with $v_a$ 1rows, $W_b$ should be a $(v_a + v_b) \times n_b$ size matrix with $v_b$ 1rows and the 1row positions for these matrices should not overlap. Let $W = [W_a | W_b]$ and $L = [L_a | L_b]$. Then, $W$ should have $v_a + v_b$ unique 1rows. Looking at the relations obtained from our model, we see that $W$ is generated by receiver matrix $T$ operating on $L$, and the rows of

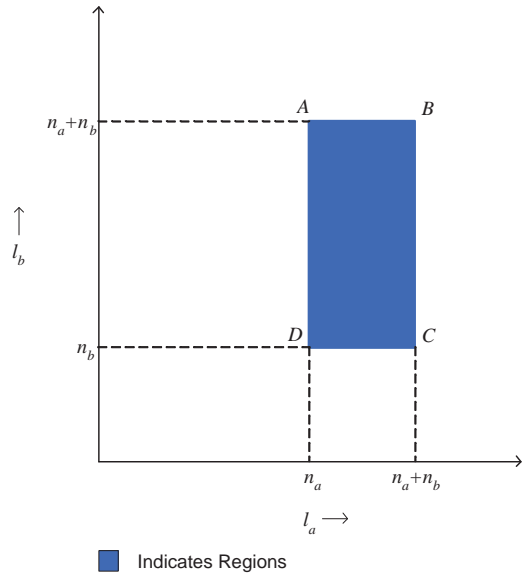

Figure 13: Region of Analysis.

$W$ are linear combinations of the rows of $L$. By definition, $W$ consists only of 1rows. For a given $L$, we need to find the maximal number of 1rows in $W$ that can be generated by linear combinations of the rows of $L$. This maximizes $v_a + v_b$ for a given $l_a + l_b$ which in turn maximizes the code rate and also specifies $L_a$, $L_b$ and $T$. Therefore, codes that achieve the maximal code rate and sum rate are found by jointly optimizing the encoder and decoder matrices. In our further discussion, $I_{k \times k}$ represents a $k \times k$ identity matrix and $0_{p_1 \times p_2}$ a $p_1 \times p_2$ null matrix. In this subsection (A.4.1), we will prove Lemmas (9-12). Note that the scope of these lemmas are limited to the particular systematic code construction we are considering in this subsection. We start by proving the following lemma:

**Lemma 9** *Let $B_1$ and $B_2$ be diagonal square matrices of size $n_b$ with all diagonal elements non-zero. If $s \in [0, n_a]$ unique 1rows are added into matrix $J = \begin{bmatrix} B_1 0_{n_b \times (n_a - n_b)} B_2 \end{bmatrix}$ to obtain matrix $L$ such that the number of independent row vectors in $L$ is $s + n_b$ and $k \in [0, n_a - n_b]$ 1rows have their non-zero element in $[n_b + 1, n_a]$, then the maximal number of 1rows possible by any linear combination of the rows of $L$ is $2s - k$.*

*Proof:* The inserted 1rows that are non-zero in positions $[n_b + 1, n_a]$ cannot give rise to any other 1row in $L$ since all the rows of $J$ are 0 in that position. Any other inserted 1row can be combined with a row vector in $J$ to give a unique 1row in $L$ since all the row vectors of $L$ are independent. Thus, the $s - k$ distinct 1rows whose non-zero elements are not in the interval $[n_b + 1, n_a]$, can generate $2(s - k)$ 1rows in $L$. The total number of 1rows that can be generated by the $s$ inserted 1rows is $2(s - k) + k = 2s - k$.   □

Note that the proof depends only on the constraint that $B_1$ and $B_2$ are diagonal with non-zero diagonal elements. Hence, we set $B_1 = B_2 = I_{n_b \times n_b}$. Lemmas 10 and 11 exclude certain regions of rectangle $ABCD$ in Figure 13 since optimal codes do not exist over them.

**Lemma 10** *Optimal codes are not contained in the region $n_b < l_b \leq n_a$.*

*Proof:* Let $P = \begin{bmatrix} I_{n_a \times n_a} & | & \begin{matrix} I_{n_b \times n_b} \\ 0_{(n_a - n_b) \times n_b} \end{matrix} \end{bmatrix}$. We form matrix $L$ by inserting distinct 1rows into $P$. Adding $m_a$ bits of redundancy at transmitter a generates $m_a$ 1rows in $L$. These 1rows have their 1s in the first $n_a$ positions. Thus, if further increase in the number of 1rows in $L$ is desired by adding redundant bits at transmitter b, then the size of the redundancy added at transmitter b must satisfy

$$\begin{aligned} m_b &> m_a + n_a - n_b \\ &> n_a - n_b. \end{aligned}$$

Thus $m_b = 0$ (no redundancy added at transmitter b) or $m_b > n_a - n_b$, which completes the proof. □

**Lemma 11** *Codes that achieve the maximal code rate and capacity are not contained on the line $l_a = l_b$.*

*Proof:* Let $P$ be defined as before. Coding on this line results in the insertion of at least one row vector to $P$ that is not a 1row. The inserted rows that are not 1rows contain one 1 in the first $n_a$ positions and one 1 in the last $n_b$ positions. The other elements are 0. The number of 1rows determines the size of the subset recoverable at the receiver. In this case, the rows that are not 1rows increase redundancy but do not increase the number of information bits recoverable at the receiver. Thus, we should not insert any row that is not a 1row. This is not possible on the line $m_a = m_b - [n_a - n_b]$. Hence, this line does not contain optimal codes and the statement of theorem follows. □

**Structure of Generator Matrices**

We now develop the structure of the generator matrices for the encoders at the two transmitters. We use *systematic codes* and show that they are optimal in terms of achieving maximal code rate and capacity. Using the results of lemmas 10 and 11, we now reduce the rectangular region $ABCD$ in Figure 13 into three sub-regions shown in Figure 14. Let us first establish two cases:

**Case 1**: $m_b = 0$.

In this case, redundancy is added at transmitter a only. Let the redundancy added to $\vec{a}$ be $m_a$. This corresponds to appending $m_a$ 1row vectors to $P$ such that the 1 in each of these vectors lies in the first $n_a$ positions and the resulting matrix consists of independent rows. Using Lemma 9, the maximal number of 1rows that can be generated is $2(m_a + n_a - n_b) - (n_a - n_b) = 2m_a + n_a - n_b$. Now, $[n_a - n_b] + m_a$ of the 1rows generated will have their 1 in the first $n_a$ positions and $m_a$ 1rows will have their 1 in the last $n_b$ position. Thus we have:

$$v_a = [n_a - n_b] + m_a \qquad v_b = m_a,$$

$$L_a = \begin{bmatrix} I_{n_a \times n_a} \\ M_{m_a \times n_a} \\ 0_{(n_b - m_a) \times n_a} \end{bmatrix} \qquad L_b = \begin{bmatrix} I_{n_b \times n_b} \\ 0_{n_a \times n_b} \end{bmatrix},$$

where, $m_b = 0$, $0 \leq m_a \leq n_a$ and $M$ is a matrix containing 1rows.

**Case 2** : $n_a - n_b < m_b \leq n_a$.

Let $m_b = n_a - n_b + k$, where $0 < k \leq n_b$.

When $m_a < k$, $m_a$ 1rows are appended to $P$ such that each 1row contains a 1 in the first $n_a$ positions. Then, $k - m_a$ 1rows are appended to the matrix resulting from the previous step so that a 1 is contained in one of the last $n_b$ positions. The 1rows are appended such that all rows of the resulting matrix are independent. Using Lemma 9, we see that the maximal number of 1rows that can be generated is given by $2m_b - [n_a - n_b]$. There are $m_b$ 1rows with 1 in the first $n_a$ positions and $m_b - [n_a - n_b]$ 1rows with 1 in the last $n_b$ positions. Thus we have:

$$v_a = m_b \qquad v_b = m_b - [n_a - n_b],$$

$$L_a = \begin{bmatrix} I_{n_a \times n_a} \\ \Lambda_{m_a \times n_a} \\ 0_{(n_b - m_a) \times a} \end{bmatrix} \qquad L_b = \begin{bmatrix} I_{n_b \times n_b} \\ 0_{(n_a - n_b + m_a) \times n_b} \\ \Delta_{(m_b - m_a - [n_a - n_b]) \times n_b} \\ 0_{(n_a - m_b) \times n_b} \end{bmatrix},$$

where, $[n_a - n_b] < m_b \leq n_a$ and $0 \leq m_a < m_b - [n_a - n_b]$. $\Lambda$ and $\Delta$ are matrices containing unique 1rows.

When $m_a > k$, coding involves appending $k$ 1rows to $P$ such that each row contains the 1 in the last $n_b$ positions. Then, $m_a - k$ 1rows are appended to the matrix resulting from the previous step so that a 1 is contained in the first $n_a$ positions for each vector. 1rows are appended such that the resulting matrix consists of independent rows. Using Lemma 9, we see that the maximum number of 1rows that can be generated is given by $2m_a + [n_a - n_b]$. There are $m_a$ 1rows with 1 in the last $n_b$ positions and $m_a + [n_a - n_b]$ 1rows with 1 in the first $n_a$ positions. Thus, we have:

$$v_a = m_a + [n_a - n_b] \qquad v_b = m_a,$$

$$L_a = \begin{bmatrix} I_{n_a \times n_a} \\ 0_{(m_b - [n_a - n_b]) \times n_a} \\ S_{(m_a - m_b + [n_a - n_b]) \times n_a} \\ 0_{(n_b - m_a) \times n_a} \end{bmatrix} \qquad L_b = \begin{bmatrix} I_{n_b \times n_b} \\ 0_{(n_a - n_b) \times n_b} \\ K_{(m_b - [n_a - n_b]) \times n_b} \\ 0_{(n_b - m_b + n_a - n_b) \times n_b} \end{bmatrix},$$

where, $[n_a - n_b] < m_b \leq n_a$ and $0 \leq m_a < m_b - [n_a - n_b]$. $S$ and $K$ are matrices containing 1rows.

**Regions**

The regions over which optimal codes exist can be described and are shown in Figure 14. We denote the code rate in regions 1, 2 and 3 as $C_{rate-R_1}$, $C_{rate-R_2}$ and $C_{rate-R_3}$, respectively.

Region 1: $n_b \leq l_a \leq n_a + n_b$ and $l_b = n_b$,

$$v_a = m_a + [n_a - n_b] \qquad v_b = m_a \qquad C_{rate-R_1} = \frac{2m_a + [n_a - n_b]}{n_a + n_b + m_a}.$$

Region 2: $l_b < l_a \leq n_a + n_b$ and $n_a < l_b \leq n_a + n_b$,

$$v_a = m_a + [n_a - n_b] \qquad v_b = m_a \qquad C_{rate-R_2} = \frac{2m_a + [n_a - n_b]}{n_a + n_b + m_a + m_b}.$$

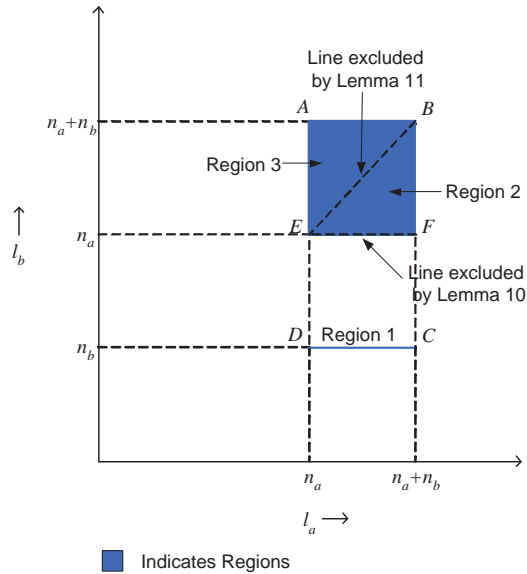

Figure 14: Gross (un-optimized) regions over which optimal codes exist.

Region 3: $n_a \leq l_a < l_b$ and $n_a < l_b \leq n_a + n_b$,

$$v_a = m_b \qquad v_b = m_b - [n_a - n_b] \qquad C_{rate-R_3} = \frac{2m_b - [n_a - n_b]}{n_a + n_b + m_a + m_b}.$$

**Optimized Regions**

**Lemma 12** *To achieve the maximum code rate, it suffices to add redundancy to only one vector.*

*Proof :* We see from Figure 14 that in Region 1 and Region 2, $v_a$ and $v_b$ do not depend upon $m_b$. Thus, for higher code rate, $m_b$ should be kept as low as possible. We thus set $m_b = 0$ for Region 1 and $m_b = n_a - n_b + 1$ for Region 2. As $n_a \geq n_b$,

$$C_{rate-R_1} \geq C_{rate-R_2}.$$

Thus optimal codes cannot be in Region 2, as this region does not contain codes with higher code rate than Region 1. Hence, we do not consider this region in our further search for optimal codes. In Region 3, $v_a$ and $v_b$ do not depend on $m_a$. Therefore, it is best to keep $m_a$ at its lowest, i.e. $m_a = 0$. We thus consider codes over Region 1 and Region 3, where we set $m_b = 0$ and $m_a = 0$, respectively. Thus, in order to achieve the optimal code rate, it suffices to add redundancy at only one transmitter. □

The optimized regions are shown in Figure 15. The code rate in regions A and B are denoted as $C_{rate-R_A}$ and $C_{rate-R_B}$, respectively.

Region A : $n_a \leq l_a \leq n_a + n_b$ and $l_b = n_b$

$$v_a = m_a + [n_a - n_b] \qquad v_b = m_a \qquad C_{rate-R_A} = \frac{2m_a + [n_a - n_b]}{n_a + n_b + m_a}.$$

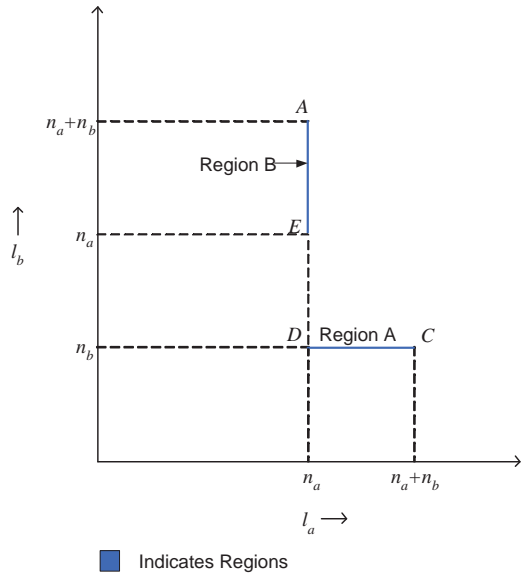

Figure 15: Optimized regions.

Region B: $l_a = n_a$ and $n_a + 1 \leq l_b \leq n_a + n_b$

$$v_a = m_b \qquad v_b = m_b - [n_a - n_b] \qquad C_{rate-R_B} = \frac{2m_b - [n_a - n_b]}{n_a + n_b + m_b}.$$

**Achieving the capacity region and maximal code rate**

Lemma 12 shows that in order to achieve the maximal code rate, it suffices to add redundancy at only one transmitter. Let the redundancy be $m$. In Region A, $m_a = m$, $0 \leq m \leq n_b$ and

$$C_{rate-R_A} = \frac{2m + [n_a - n_b]}{n_a + n_b + m}.$$

In Region B, $m_b = m$, $(n_a - n_b + 1) \leq m \leq n_b$ and

$$C_{rate-R_B} = \frac{2m - [n_a - n_b]}{n_a + n_b + m}.$$

**Case 1 :** $n_a > n_b$. When $0 \leq m \leq n_a - n_b$, Region B is excluded and Region A provides the only solution. For all other $m$, $C_{rate-R_A} > C_{rate-R_B}$. Thus, Region A always provides a higher code rate than Region B. From the code rate equations derived earlier, we see that the maximal code rate is obtained when $m$ is largest, i.e. $m = n_b$. Therefore, $(m_a, m_b) = (n_b, 0)$ is the optimal point. This corresponds to $(l_a, l_b) = (n_a + n_b, n_b)$. Thus, for obtaining the maximal code rate, we add redundancy to only the larger transmit vector and the size of the redundancy is the size of the smaller transmit vector. The code rate is

$$C_{rate} = \frac{n_a + n_b}{n_a + 2n_b}. \tag{48}$$

**Case 2:** $n_a = n_b = n$. Here, for a given $m$, both regions give the same code rate and we can add redundancy to either of the two vectors. A symmetry exists about the line $l_a = l_b$ and there are two optimal points. Code rate is maximal when $m$ is maximal, i.e. $m = n$. These points are $(m_a, m_b) \in \{(0, n), (n, 0)\}$ corresponding to $(l_a, l_b) \in \{(2n, n), (n, 2n)\}$. In this case, coding results in the size of redundancy being equal to the transmit vector size and the code rate is 2/3.

The transmission rates of the code are

$$R_a = \frac{n_a}{n_a + n_b}, \tag{49}$$

$$R_b = \frac{n_b}{n_a + n_b}, \tag{50}$$

$$R_{sum}^{NSMA} = R_a + R_b = 1. \tag{51}$$

We see from (48,51) that this code achieves the maximum code rate and capacity for this channel and is thus an optimal code. Moreover, this code obeys the property that no redundancy is added to the smaller transmit vector. Theorem 16 proves that any maximal code rate achieving code satisfies this property.

### A.4.2

**Proof of Theorem 16:**

We first prove the forward part. Let a code be capacity approaching without redundancy being added to the smaller transmit vector. From the code construction described in Section 4.5, we see that such codes exist. We thus have the following relations:

$$\begin{aligned} S &= n_a + n_b, \\ n_b &= \min(l_a, l_b), \\ C_{rate} &= \frac{n_a + n_b}{\min(l_a, l_b) + S}, \\ \Rightarrow C_{rate} &= \frac{n_a + n_b}{n_a + 2n_b}. \end{aligned}$$

Therefore, capacity approaching codes with no redundancy added to the smaller transmit vector achieve the maximal code rate by (17). This completes the forward part of the proof. We prove the reverse part now. A code that achieves the maximal code rate must meet the inequality in (16) with equality. From the code construction described in Section 4.5, we know that codes that achieve the maximal code rate exist. Therefore, maximal code rate achieving codes must satisfy

$$\begin{aligned} S &= n_a + n_b, \\ \min(l_a, l_b) &= n_b. \end{aligned}$$

Hence, maximal code rate achieving codes achieve capacity and do not add redundancy to the smaller transmit vector. The reverse part of the proof is now complete and we have

proved the theorem. □

**Proof of Theorem 17**
Let the probability that transmitters a and b have a codeword to transmit in a slot be $p_a$ and $p_b$ respectively. Transmissions begin at the start of a slot and the transmitters do not coordinate. We again define the sizes of $\vec{a}$ and $\vec{b}$ as $n_a$ and $n_b$. Consider the *expected code rate*. The mean sizes of $\vec{a}$ and $\vec{b}$ are $p_a n_a$ and $p_b n_b$ respectively. We will therefore define the mean sizes $n'_1, n'_2$ where $n'_2 \leq n'_1$ as

$$n'_1 = \max(p_a n_a, p_b n_b) \qquad n'_2 = \min(p_a n_a, p_b n_b).$$

The maximal expected code rate is thus

$$C_{rate} = \frac{n'_1 + n'_2}{n'_1 + 2n'_2},$$

where (from Theorem 16) no redundancy is added to the transmit vector with smaller mean size. Therefore, two cases arise.

**Case 1:** We add redundancy only at a and not at b if the mean size of $\vec{a}$ is greater than the mean size of $\vec{b}$. This implies that

$$n'_1 = p_a n_a \qquad n'_2 = p_b n_b,$$
$$C_{rate} = \frac{p_a n_a + p_b n_b}{p_a n_a + 2 p_b n_b}.$$

**Case 2:** We add redundancy only at b and not at a if the mean size of $\vec{a}$ is smaller than the mean size of $\vec{b}$, which implies that

$$n'_1 = p_b n_b \qquad n'_2 = p_a n_a,$$
$$C_{rate} = \frac{p_a n_a + p_b n_b}{2 p_a n_a + p_b n_b}.$$

When $n'_1 = n'_2$, we can use either technique. Figure 16 shows the regions of the two dimensional space of $(p_a, p_b)$ where the cases apply. Region 1 corresponds to the first case and Region 2 to the second.

Let us denote $\alpha = \frac{p_a}{p_b}$ and $\beta = \frac{n_a}{n_b}$. We have $\alpha \in [0, \infty)$ and $\beta \geq 1$. Thus, the maximal expected code rate expression can be written as

$$C_{rate}(\alpha, \beta) = \frac{1 + \alpha\beta}{1 + \alpha\beta + \min(1, \alpha\beta)}.$$

For $\alpha \in [0, \frac{1}{\beta}]$, the mean size of $\vec{a}$ is less than or equal to the size of $\vec{b}$ and we add redundancy only at b. For $\alpha \in [\frac{1}{\beta}, \infty)$ the mean size of $\vec{a}$ is larger than or equal to the mean size of $\vec{b}$, and we add redundancy only at a. Note that for $\alpha = \frac{1}{\beta}$, we may add redundancy at a or b and still obtain the same expected code rate. The expected code rate is minimal and has a value of 2/3 when $p_a n_a = p_b n_b$. Figure 17 shows how the expected code rate changes with $\alpha$. We now look at the limit when transmitter a stops transmitting, i.e. $\alpha \to 0$ and when transmitter b stops transmitting, i.e. $\alpha \to \infty$. Evaluating the value of expected code rate as $\alpha$ tends to 0 or $\infty$, we get

$$\lim_{\alpha \to 0} C_{rate}(\alpha, \beta) = 1 \qquad \lim_{\alpha \to \infty} C_{rate}(\alpha, \beta) = 1.$$

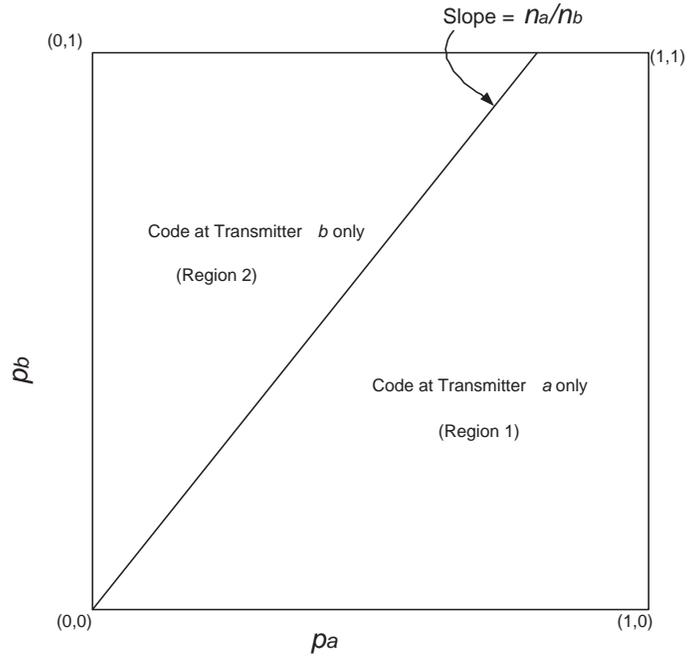

Figure 16: Coding Regions for Bursty Multiple Access.

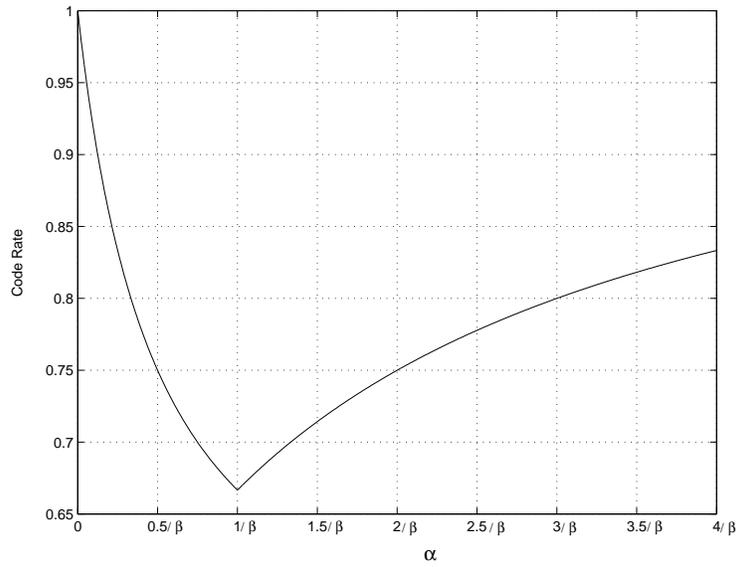

Figure 17: Variation of code rate with $\alpha$.

These limits are what we had expected since, in both cases, one transmitter transmits in a slot with probability 1 and the other does not transmit at all. There is no multiple access interference and the average code rate becomes the code rate of a point-to-point noiseless channel, i.e. 1.

When $\beta = 1$, i.e. $n_a = n_b$, we see that if $p_b \leq p_a$, we add redundancy only at a and when $p_b > p_a$ we add redundancy only at b. The statement of the theorem follows. $\square$

# Appendix 5

**Proof of Theorem 18:**
In this case, the linear encoder is a matrix of dimension

$$(\lceil nR_1 \rceil + \lceil nR_2 \rceil + \lceil nR_3 \rceil + \lceil nR_{12} \rceil + \lceil nR_{23} \rceil + \lceil nR_{13} \rceil + \lceil nR_{123} \rceil) \times n.$$

The first $\lceil nR_1 \rceil$ bits of the output go to receiver 1 only. The subsequent $\lceil nR_2 \rceil$ and $\lceil nR_3 \rceil$ bits similarly go to receivers 2 and 3, respectively. Next come, in order, the rate-$R_{12}$, $R_{23}$, $R_{13}$, and $R_{123}$ descriptions. We again use typical set decoding.

Given the linear structure of the code, we can break encoder matrix $A_n$ into a collection of $\lceil nR_a \rceil \times n$ sub-matrices, $a \in \{1, 2, 3, 12, 23, 13, 123\}$, such that

$$A_n = \begin{bmatrix} A_{1,n} \\ A_{2,n} \\ A_{3,n} \\ A_{12,n} \\ A_{23,n} \\ A_{13,n} \\ A_{123,n} \end{bmatrix}.$$

We begin by bounding the expected probability of decoding in error at receiver 1, here denoted as $E[P_e(A_{1,n}, A_{12,n}, A_{13,n}, A_{123,n})]$. The arguments for receivers 2 and 3 are similar. By the union bound, the code error probability is bounded by the sum of the individual decoder error probabilities.

An error occurs at receiver 1 if any subset of the desired sources is decoded in error. Thus, following our standard approach,

$$\begin{aligned} E[P_e(A_{1,n}, A_{12,n}, A_{13,n}, A_{123,n})] &\leq \epsilon_n + \sum_{(u_1^n, u_{12}^n, u_{13}^n, u_{123}^n) \in A_\epsilon^{(n)}} p(u_1^n, u_{12}^n, u_{13}^n, u_{123}^n) \\ &\quad \cdot \sum_{s \subseteq S_1 : s \neq \phi} \sum_{\hat{u}_s^n \neq u_s^n : (\hat{u}_s^n, u_{S_1-s}^n) \in A_\epsilon^{(n)}} \Pr(A_{s,n}(\mathbf{u}_s - \hat{\mathbf{u}}_s) = \mathbf{0}) \\ &\leq \epsilon_n + \sum_{s \subseteq S_1 : s \neq \phi} 2^{n(H(U_s | U_{S_1-s}) + 2\epsilon)} 2^{-(nR)_s} \end{aligned}$$

for some $\epsilon_n \to 0$. □

**Proof of Lemma 5:**
By [1, Theorem 14.6.1,Theorem 14.6.2], the capacity of the given channel is the convex hull of the closure of all $(R_1, R_2)$ satisfying $R_2 \leq I(W; Y_2)$ and $R_1 \leq I(X; Y_1|W)$ for some joint distribution $p(w)p(x|w)p(y_1|x)p(y_2|y_1)$. Here $W$ is an auxiliary random variable with alphabet size 2 and $p(y_2|y_1)$ is derived from the physically degraded channel model. By a symmetry argument, the optimal $W$ is a uniform binary random variable with $p(x|w) = 1-\beta$ if $x = w$ and $p(x|w) = \beta$ otherwise. Thus

$$\begin{aligned} R_1 &\leq I(X; Y_1|W) \\ &= I(X; Y_1) - I(W; Y_1) \\ &= (1 - q_1(1)) - [H((1 - q_1(1))/2, q_1(1), (1 - q_1(1))/2) \\ &\quad - H((1 - \beta)(1 - q_1(1)), q_1(1), \beta(1 - q_1(1)))] \\ &= (1 - q_1(1))H(\beta) \\ R_2 &\leq I(W; Y_2) \\ &= H((1 - q_1(1))(1 - \alpha)/2, q_1(1) + (1 - q_1(1))\alpha, (1 - q_1(1))(1 - \alpha)/2) \\ &\quad - H((1 - \beta)(1 - q_1(1))(1 - \alpha), q_1(1) + (1 - q_1(1))\alpha, \beta(1 - q_1(1))(1 - \alpha)) \\ &= (1 - q_1(1))(1 - \alpha)(1 - H(\beta)) \\ &= (1 - q_2(1))(1 - H(\beta)). \end{aligned}$$

Varying $H(\beta)$ from 0 to 1 gives the independent coding result. The common information result comes from [1, Theorem14.6.4]. □

**Strategy for moving linear codes linear codes beyond the time-sharing bound:**
Consider a systematic code with a low density parity-check matrix. Let the encoding matrix be

$$B_n = \begin{bmatrix} I \\ P_{11} & P_{21} \\ 0 & P_{22} \end{bmatrix},$$

where $I$ is the $(\lfloor nR_1 \rfloor + \lfloor nR_2 \rfloor) \times (\lfloor nR_1 \rfloor + \lfloor nR_2 \rfloor)$ identity matrix and $P_{11}$, $P_{21}$, and $P_{22}$ have dimensions

$$nR_1 \frac{H(Z_1)}{1 - H(Z_1)} \times nR_1, \quad nR_1 \frac{H(Z_1)}{1 - H(Z_1)} \times nR_2, \quad \left(n - nR_2 - nR_1 \frac{1}{1 - H(Z_1)}\right) \times nR_2,$$

respectively. (We here drop the rounding notation for readability but note that all of the above quantities must be integers.) For each $i \in \{1, 2\}$, let $\mathbf{Z}_i^t = [\mathbf{Z}_{i1}^t \mathbf{Z}_{i2}^t \mathbf{Z}_{i3}^t \mathbf{Z}_{i4}^t]$, where the sub-vectors have lengths $nR_1$, $nR_2$, $nR_1 H(Z_1)/(1 - H(Z_1))$, and $n - nR_2 - nR_1/(1 - H(Z_1))$, respectively. Applying the above code, the channel output at receiver 2 is

$$\mathbf{Y}_2 = \begin{bmatrix} \mathbf{V}_1 + \mathbf{Z}_{21} \\ \mathbf{V}_2 + \mathbf{Z}_{22} \\ P_{11}\mathbf{V}_1 + P_{21}\mathbf{V}_2 + \mathbf{Z}_{23} \\ P_{22}\mathbf{V}_2 + \mathbf{Z}_{24} \end{bmatrix}.$$

If the decoder at that receiver applies parity check matrix $P_{11}$ to the the received $\mathbf{V}_1 + \mathbf{Z}_{21}$ and subtracts off the outcome from the third component of $\mathbf{Y}$ then the modified signal is

$$\mathbf{Y}_2 - \begin{bmatrix} \mathbf{0} \\ \mathbf{0} \\ P_{11}(\mathbf{V}_1 + \mathbf{Z}_{21}) \\ \mathbf{0} \end{bmatrix} = \begin{bmatrix} \mathbf{V}_1 + \mathbf{Z}_{21} \\ \mathbf{V}_2 + \mathbf{Z}_{22} \\ P_{21}\mathbf{V}_2 + \mathbf{Z}_{23} + P_{11}\mathbf{Z}_{21} \\ P_{22}\mathbf{V}_2 + \mathbf{Z}_{24} \end{bmatrix}.$$

Decoder 2 thereby recovers more of its parity check symbols at the expense of increasing the corresponding error probability in those symbols. When the density of parity check matrix $P_{11}$ is low, the increase in error probability for symbols $P_{21}\mathbf{V}_2$ may also be low enough to make those parity check bits useful in decoding the description of $\mathbf{V}_2$. Receiver 1 uses the same technique to decode $\mathbf{V}_2$, then subtracts off its impact on the parity check bits for $\mathbf{V}_1$, and finally decodes $\mathbf{V}_1$.